\documentclass[]{spie}  

\usepackage{amsmath,amsfonts,amssymb}
\usepackage{graphicx}
\usepackage[colorlinks=true, allcolors=blue]{hyperref}
\usepackage{orcidlink}
\usepackage{booktabs} 
\usepackage{multirow}
\usepackage{enumitem}
\usepackage[labelfont=bf]{caption}
\usepackage{subcaption}

\usepackage{siunitx}
\DeclareSIUnit\dBm{dBm}
\DeclareSIUnit{\evt}{evt\per\second}
\DeclareSIUnit{\kevt}{\kilo\evt}
\DeclareSIUnit{\rpm}{rpm}
\DeclareSIUnit{\px}{pixels}
\DeclareSIUnit{\radiant}{rad}
\DeclareSIUnit{\counts}{counts}
\DeclareSIUnit{\usd}{USD}
\usepackage{algorithm}
\usepackage{algpseudocode}

\usepackage{amsmath}

\usepackage{tikz}
\usetikzlibrary{shapes.geometric, positioning, arrows.meta, fit, calc, patterns}
\usepackage{anyfontsize}
\usepackage{pgfplots}
\pgfplotsset{compat=1.18}

\usepackage{eso-pic}
\usepackage{tcolorbox}
\AddToShipoutPicture*{%
  \AtPageUpperLeft{%
    \raisebox{-1.5cm}{ 
      \makebox[\paperwidth]{%
        \centering
        \begin{minipage}{1\textwidth} 
          \centering
          \scriptsize 
          This paper has been accepted for invited presentation at 
          SPIE Sensors + Imaging 2026, Edinburgh, UK. ©2026 SPIE. Please cite it as: \\
          \textit{Francesco Nardo, Nigitha Selvaraj, Klaus Schertler, et al. "Neuromorphic camera for multitarget pointing acquisition and tracking in free-space optical communication", Proc. SPIE 14281, Environmental Effects on Light Propagation and Adaptive Systems IX, 142810A (10 Nov 2026); \url{https://doi.org/10.1117/12.3116799}}
        \end{minipage}
      }
    }
  }
}
\AddToShipoutPicture*{ 
  \AtPageLowerLeft{%
    \raisebox{0.3cm}{ 
      \makebox[\paperwidth]{%
        \begin{tcolorbox}[colframe=black, colback=white, sharp corners, width=\textwidth, ]
          \scriptsize
          ©2026 Society of Photo‑Optical Instrumentation Engineers (SPIE). One print or electronic copy may be made for personal use only. Systematic reproduction and distribution, duplication of any material in this publication for a fee or for commercial purposes, and modification of the contents of the publication are prohibited.
        \end{tcolorbox}
      }
    }
  }
}

\title{Neuromorphic Camera for Multi-Target Pointing Acquisition and Tracking in Free Space Optical Communication}

\author[a,b]{Francesco Nardo \orcidlink{0009-0006-5750-5837} }
\author[a,c]{Nigitha Selvaraj}
\author[a]{Klaus Schertler}
\author[a]{Peter Krötz}
\author[a]{Ricardo Barrios \orcidlink{0000-0002-2545-5568} }
\author[c]{Sebastian Houben \orcidlink{0000-0002-2036-419X} }
\author[b]{Sebastian Randel}
\affil[a]{Airbus Defence and Space GmbH, Taufkirchen, Germany}
\affil[b]{Institute of Photonics and Quantum Electronics, Karlsruhe Institute of Technology, Germany}
\affil[c]{Institute for Artificial Intelligence and Autonomous Systems, University of Applied Sciences Bonn-Rhein-Sieg, Germany}

\authorinfo{Further author information: (Send correspondence to F.Nardo)\\F.Nardo: E-mail: fn.francesconardo@gmail.com}

\begin{document} 
\maketitle

\begin{abstract}
Pointing, Acquisition, and Tracking (PAT) in Free-Space Optical Communication (FSOC) applications must achieve both low jitter and a large Field of View (FoV) to support beacon acquisition and multi-target tracking. While conventional frame-based sensors provide the necessary angular coverage, their performance is limited by latency, available frame-rate, and motion blur. This work explores and demonstrates the application of an event-based vision sensor for PAT tasks in FSOC.
The system employs a neuromorphic sensor that asynchronously records brightness changes (events) with microsecond-level temporal resolution. The main contribution of this paper is a comparative analysis of event-based tracking algorithms for the simultaneous acquisition and tracking of multiple optical beacons that operate directly on this asynchronous stream. The performance of these strategies is validated on an experimental dataset representing realistic operational conditions for dynamic High-Altitude Platform (HAP) and Unmanned Aerial Vehicle (UAV) networks. The analysis compares key metrics, such as tracking accuracy and processing latency, under various signal-to-background conditions.
\end{abstract}


\section{Introduction}\label{sec:intro} 

Free Space Optical Communication (FSOC) is a key technology for future high capacity wireless networks due to its large available bandwidth, immunity to Radio Frequency (RF) congestion, and license-free operation. 
However, compared to RF solutions, laser beams are highly directional and therefore require a very precise Pointing, Acquisition, and Tracking (PAT) system.
Considering the receiver side, the following properties will enable simplified optical communication between multiple High Altitude Platforms (HAPs) or Unmanned Aerial Vehicles (UAVs):
(i) large Field of View (FoV), 
(ii) high sensitivity, 
(iii) low latency / high tracking bandwidth and 
(iv) low cost, Size Weight and Power (SWaP). 

Currently, no single solution satisfies these requirements without compromises \cite{hemmatiNearEarthLaserCommunications2021}. 
To address these limitations, the application of Neuromorphic Vision Sensors (NVSs) or Dynamic Vision Sensors (DVSs), also known as neuromorphic cameras or event cameras, to FSOC beacon tracking is proposed~\cite{cazzatoApplicationDrivenSurveyEventBased2024}.
NVSs are large resolution sensors (millions of pixels) that asynchronously record brightness changes (events) with microsecond-level temporal resolution while featuring an extremely wide dynamic range (over \qty{120}{\dB}) \cite{lichtsteiner128$times$128120DB2008}.
Thanks to their dynamic range, and in conjunction with wide-FoV objectives, NVSs can provide faster acquisition compared to conventional gimbal receivers relying on raster or spiral scanning to locate their counter-terminals.
For instance, the $\approx \qty{6}{\minute}$ visibility window between a \qty{300}{\kilo\meter} LEO satellite and a HAP~\cite{aliPredictingVisibilityLEO1999} means that a \qty{30}{\second} acquisition latency consumes \qty{8.3}{\%} of the total link duration.

Specifically, this work investigates four distinct event-based acquisition and tracking methods for FSOC beacons, including both activity-based and frequency-based algorithms and incorporates Kalman and particle filtering to improve stability and reduce identity switching in multi-target scenarios.
The proposed algorithms are evaluated using a commercially available NVS (Sony IMX636) under realistic operating conditions, including motion, vibration, atmospheric scintillation, and signal fades. All four methods achieve a median recall above 96\%, where recall is defined as the ratio of the tracked duration to the beacon's total visibility time in the field of view. These results demonstrate that event-based sensors can reliably perform laser beacon acquisition and tracking in FSOC applications.
The fastest approach (lowest latency), i.e. the protocol-informed method, consistently has a tracking bandwidth up to \qty{870}{\hertz} with an initial beacon acquisition time $t_{\text{acq}} = \qty{20}{\milli\second}$, a latency of $\qty{650}{\micro\second}$ and a sub-pixel tracking accuracy. 
From a system design perspective, a \qty{3.5}{\mm} wide-angle objective ($\approx \qty{91}{\degree}$ FoV) with a \qty{1}{}-pixel error will have an angular accuracy of $\approx \qty{1.38}{\milli\radiant}$; whereas a \qty{100}{\mm} tele-objective ($\approx \qty{ 4.1}{\degree}$ FoV) will have an angular accuracy of $\approx \qty{48.6}{\micro\radiant}$. 

The remainder of this paper is organized as follows: \autoref{sec:art} reviews the state of the art in neuromorphic vision and event-based tracking algorithms. \autoref{sec:dataset} and \ref{sec:algorithm} explain the experimental dataset acquisition methodology and present the four proposed tracking algorithms. \autoref{sec:results} discusses the experimental results using metrics highly relevant to FSOC. Finally, \autoref{sec:conclusion} provides concluding remarks and outlines future work.

\section{State of the Art}\label{sec:art} 

In previous works, real-time object tracking has been achieved by the use of lightweight clustering approaches suitable for embedded platforms~\cite{litzenbergerEmbeddedVisionSystem2006}. 
Further work has introduced a multi-kernel approach that leverages the asynchronous nature of event streams to robustly track visual features under extreme motion conditions~\cite{lagorceAsynchronousEventBasedMultikernel2015}. 
Additionally, combining NVSs with inertial sensors and probabilistic estimators has enabled high-frequency state estimation and accurate 6-DoF tracking under dynamic motion~\cite{liAsynchronousVisualInertialOdometry2024,zhuEventBasedVisualInertial2017}.  
In multi-agent systems, event-based motion capture frameworks combine real-time object detection using deep learning models with spatial indexing structures to enable scalable multi-target tracking~\cite{iaboniEventBasedMotionCapture2022}. 

Beyond generic object tracking, NVSs have recently been explored for optical beacon detection and tracking. Existing approaches mainly rely on three core techniques: event clustering, temporal frequency analysis, and state estimation. 
First, density-based methods identify spatially and temporally correlated event clusters. For instance, real-time 2D tracking has been achieved by continuously clustering events according to their spatial proximity~\cite{litzenbergerEmbeddedVisionSystem2006}. Building on this, asynchronous blob trackers use nearest-neighbor data association combined with extended Kalman filters to estimate blob shape and motion parameters~\cite{wangAsynchronousBlobTracker2024}. 
In contrast to purely spatial clustering, frequency-based methods exploit periodic event patterns to detect optical beacons~\cite{mullerMiniatureLowpowerSensor2011}. Flickering light sources, for example, can be identified through pixel-level activity analysis before undergoing beacon verification via demodulation~\cite{wangSmartVisualBeacons2022}. Furthermore, frequency filtering effectively suppresses background activity~\cite{jianhongMethodPoseTracking2019}, while combining transition-frequency estimation with Gaussian mixture models allows for the separation of markers with distinct flicker frequencies~\cite{salahNeuromorphicVisionBasedMeasurement2024}. More recent advancements apply pixel-wise Fourier transform analysis to localize sources based on dominant frequencies~\cite{wuEMLFEventbasedFrequency2026} or accumulate on-off event intervals into frequency-specific evidence maps~\cite{censiLowlatencyLocalizationActive2013}. 
Finally, to improve tracking accuracy during extreme motion or temporary signal loss, state estimation approaches like Kalman filtering are often employed to smooth noisy detections~\cite{vonarnimDynamicEventbasedOptical2024}. Similarly, event-based optical flow estimation, including approaches driven by spiking neural networks, has been investigated to directly extract continuous motion information from the asynchronous event stream~\cite{vonarnimDynamicEventbasedOptical2024}.

In the domain of optical tracking, recent studies have demonstrated the advantages of NVSs over frame-based sensors for daylight tip-tilt star wavefront correction in astronomical adaptive optics~\cite{cockram_EventBased_2025a,cockram_Eventbased_2025b}. While these works highlight the better positional accuracy and tracking bandwidth of event-based sensors for narrow-FoV tracking from stationary ground stations, extending these benefits to wide-FoV acquisition on dynamic, vibrating platforms in FSOC networks remains an open challenge.
These methods have demonstrated successful optical marker and beacon tracking; however, most existing approaches rely on predefined blinking frequencies, simple modulation schemes, or single-target scenarios, limiting their applicability to dynamic optical communication systems with arbitrary modulation patterns and multiple transmitters.

\newpage
\section{Experimental Datasets}\label{sec:dataset}

\begin{figure}[ht]
    \centering    
    \begin{subfigure}[t]{0.437\textwidth}
        \centering
        \includegraphics[width=\linewidth]{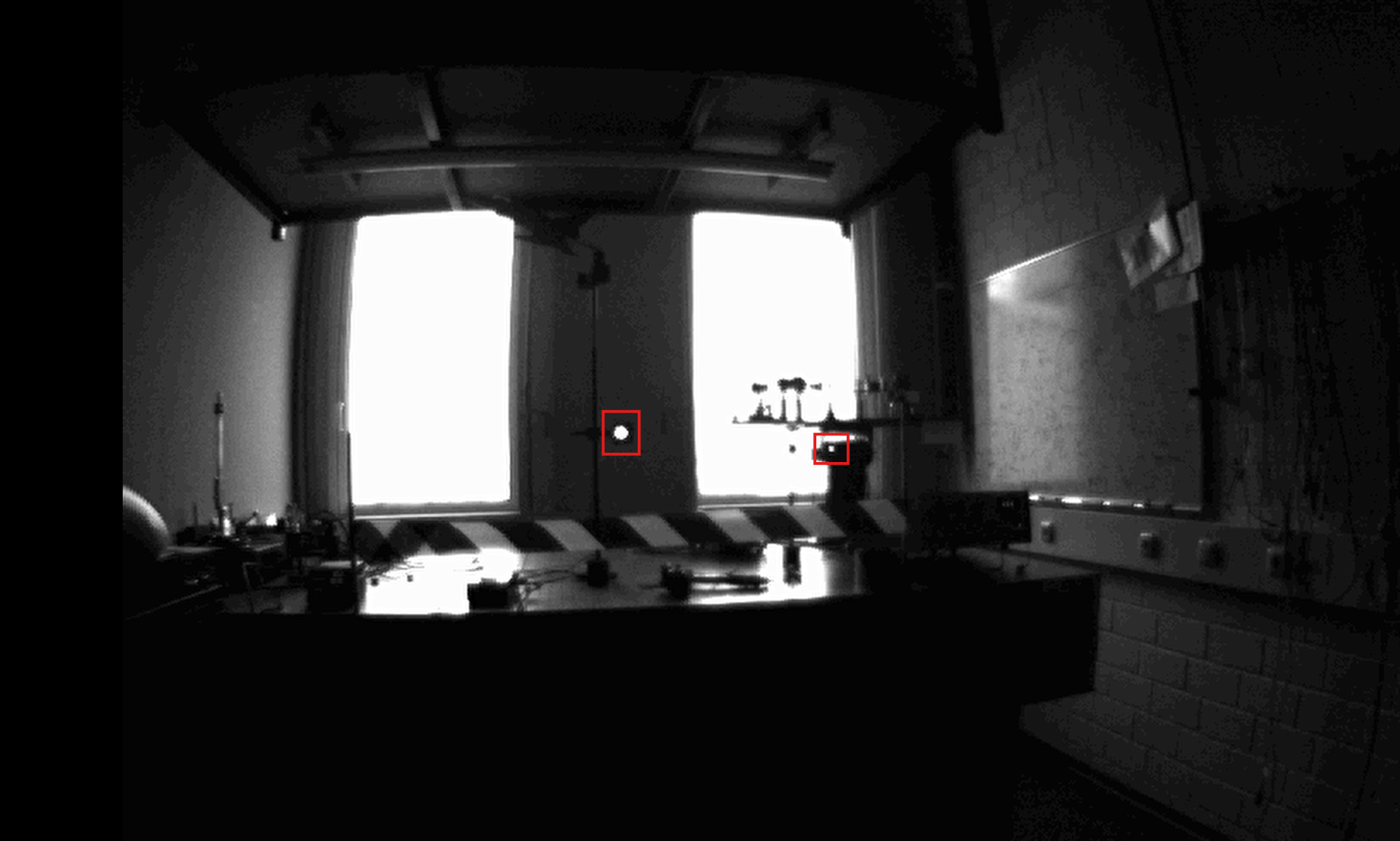}
        \label{fig:gray}
    \end{subfigure}
    \hspace{20 pt}
    \begin{subfigure}[t]{0.45\textwidth}
        \centering
        \includegraphics[width=\linewidth]{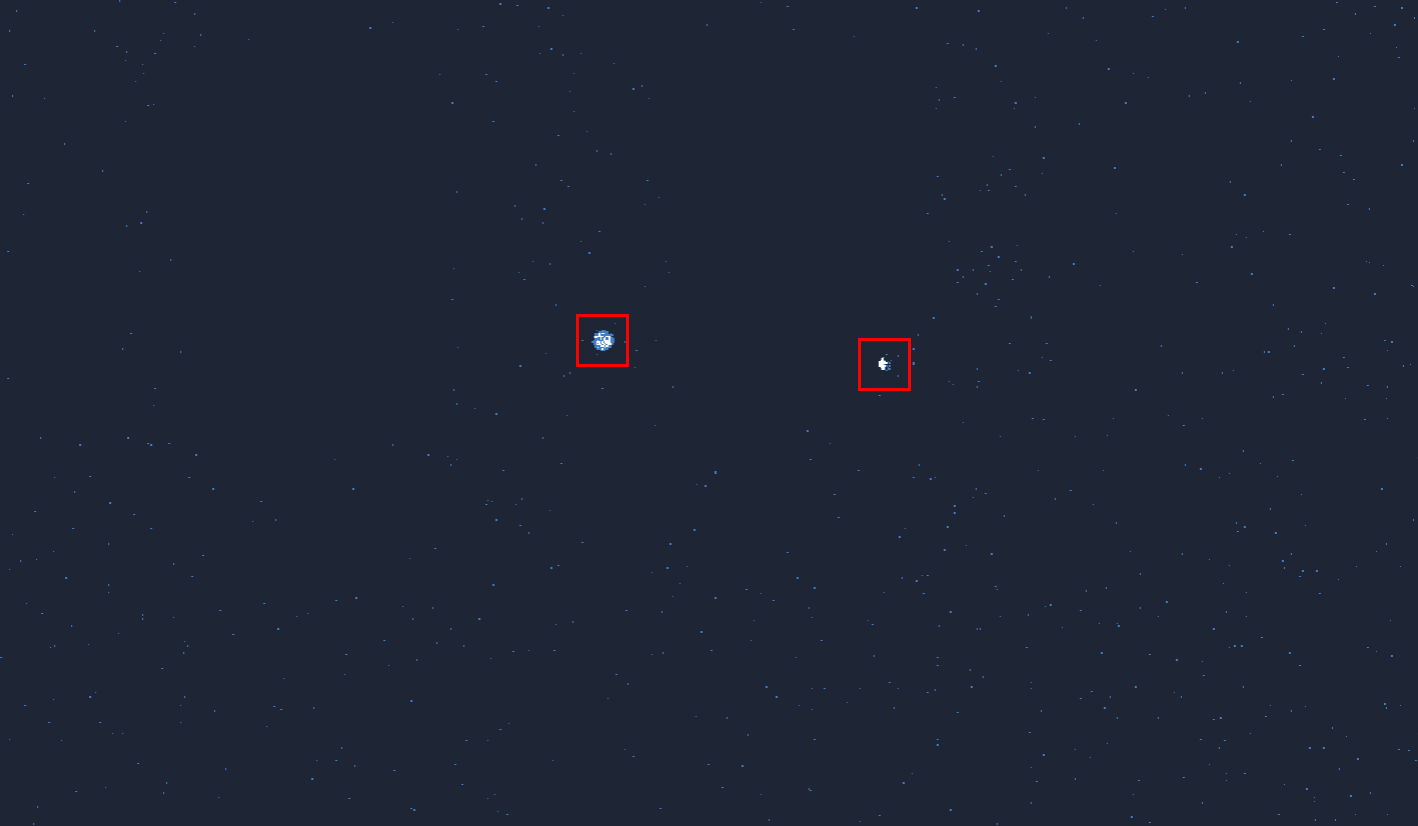}
        \label{fig:ev}
    \end{subfigure}    
    \caption{Red boxes indicate the location of the detected beacons captured by a frame-based \textbf{(Left)} and event-based \textbf{(Right)} camera in daytime with the same objective and similar resolution and pixel pitch. 
    In the gray-scale image, the left beam is clearly visible against a dark background, while the right beam is almost lost against a bright background. 
    The NVS captures only changes in light intensity, displaying higher dynamic range.}
    \label{fig:gray_ev_comp}
\end{figure}

To illustrate the wider dynamic range of neuromorphic vision sensors over frame-based cameras under daytime illumination, \autoref{fig:gray_ev_comp} provides a visual comparison of two laser beacons. 
This study uses the Prophesee EVK4 with Sony IMX636 sensor, which features a resolution of $1280 \times \qty{720}{\px}$, a \qty{4.86}{\micro\meter} pixel pitch, and a nominal temporal resolution of \qty{1}{\micro\second}. 
The sensor has spectral responsivity across \numrange{300}{1000}\qty{}{\nano\meter} (visible to near-infrared) and is operated without an optical filter, although deploying a bandpass filter matched to the beacon wavelength is recommended to suppress background noise. 
The objective is a Thorlabs MVL4WA with \qty{3.5}{\mm} effective focal length, the horizontal-FoV is \qty{83.25}{\degree}, the vertical-FoV is \qty{53.12}{\degree} and the diagonal-FoV is \qty{91.11}{\degree}.

\begin{figure}[ht]
    \centering
    \includegraphics[width=0.9\linewidth]{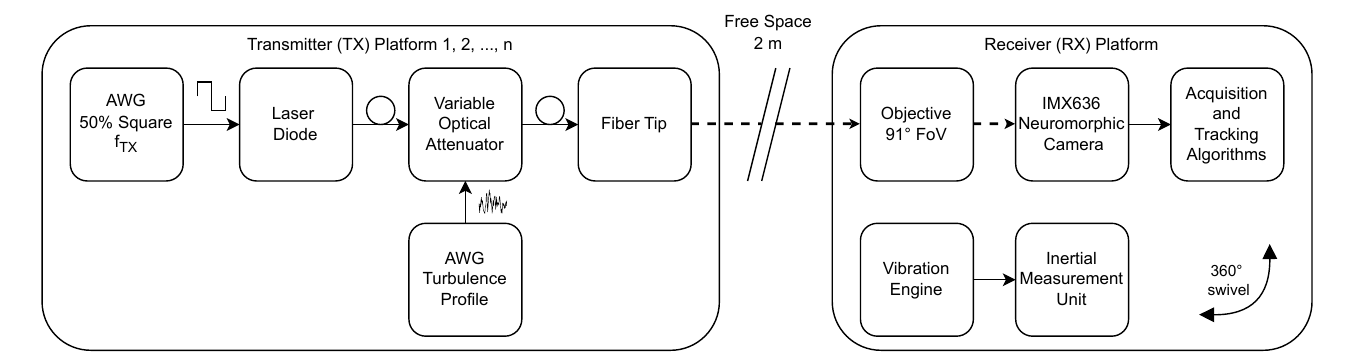}
    \caption{Diagram of the experimental setup with multiple transmitter nodes and a receiver on a motorized swivel mount. Dashed arrows indicate free-space paths, looped lines represent optical fibers, and solid arrows denote electrical analog and digital connections, with insets showing the analog signals.}\label{fig:setup-eventPAT}
\end{figure}

The experimental setup, depicted in \autoref{fig:setup-eventPAT}, comprises one Receiver (RX) and one or more Transmitter (TX) platforms.
Each TX node features an Arbitrary Waveform Generator (AWG) repeating a sequence representing a handshake or unique target identifier. 
For simplicity, the transmission is modeled as a 50\% duty-cycle square wave with variable frequency $f_{TX} = \SIrange[range-phrase=\text{ to },range-units=single]{20}{64}{\kilo\hertz}$, directly modulating a \qty{532}{\nm} or \qty{852}{\nm} laser diode.
The RX module, integrating the NVS and the Honeywell n380 Inertial Measurement Unit (IMU), is mounted on a free-floating platform supported by opposing compression springs to allow free oscillation. 
To replicate the vibration profiles experienced by aerial platforms (HAPs, UAVs, or aircrafts), ten vibration engines are installed: five PUIaudio HD-EMB1005-LW25-R (\qty{8500}{\rpm}) and five HD-EMB2110-LW140 (\qty{17000}{\rpm}). 
In flight, the vibration Power Spectral Density (PSD) of airborne FSOC terminals is dominated by propeller rotation (e.g., \SIrange{4000}{20000}{\rpm} for a LEAP-1A engine\cite{leap1A}) and engine/structural harmonics extending up to several hundred Hertz~\cite{chenDemonstrationBidirectionalCoherent2018}. 
To characterize the emulated in-flight spectra, the vibration profile is evaluated in \autoref{sec:app_vibration} using Welch's method~\cite{welchUseFastFourier1967} after applying a Hann window~\cite{harrisUseWindowsHarmonic1978} on the inertial data.
Furthermore, the RX platform is mounted on a motorized rotational stage (Newport URS50CPP) to replicate rudder yaw-axis movements, which change the beacon's angles of arrival.
Recorded conditions include constant yaw angular rates up to \qty{20}{\deg\per\s} and vibration on all three axes with up to $\approx \qty{1.7}{\m\per\s^2}$ oscillation strength.

In order to replicate the effects of atmospheric turbulence on the optical power received by the NVS, the scintillations statistics described in Andrews and Phillips \cite{andrewsLaserBeamPropagation2005} were emulated. The complete methodology is described in \autoref{sec:app_turbulence}.
The following scintillation scenarios due to turbulence are tested using a Variable Optical Absorber (VOA) (Thorlabs EVOA800) modulated by an AWG and present at the TX node:
\textbf{(i)  Atmospheric Fade:} an horizontal link between two HAPs at \qty{10}{\km} altitude and \qty{20}{\km} distance and 
\textbf{(ii) Complete Fade:} a repeated stair profile of \value{16} steps (each \qty{3}{\dB} and \qty{1}{\s} long) representing total fades (when the received power is below the NVS sensitivity). Because of the limited dynamic range of the VOA, the actual attenuation range is between \SIrange{3.1}{38.84}{\dB} however the TX power is adjusted such that at least half the stair profile is below the NVS sensitivity causing a \qty{8}{\s} long fade. 

\begin{figure}[ht]
    \centering
    \includegraphics[width=0.45\linewidth]{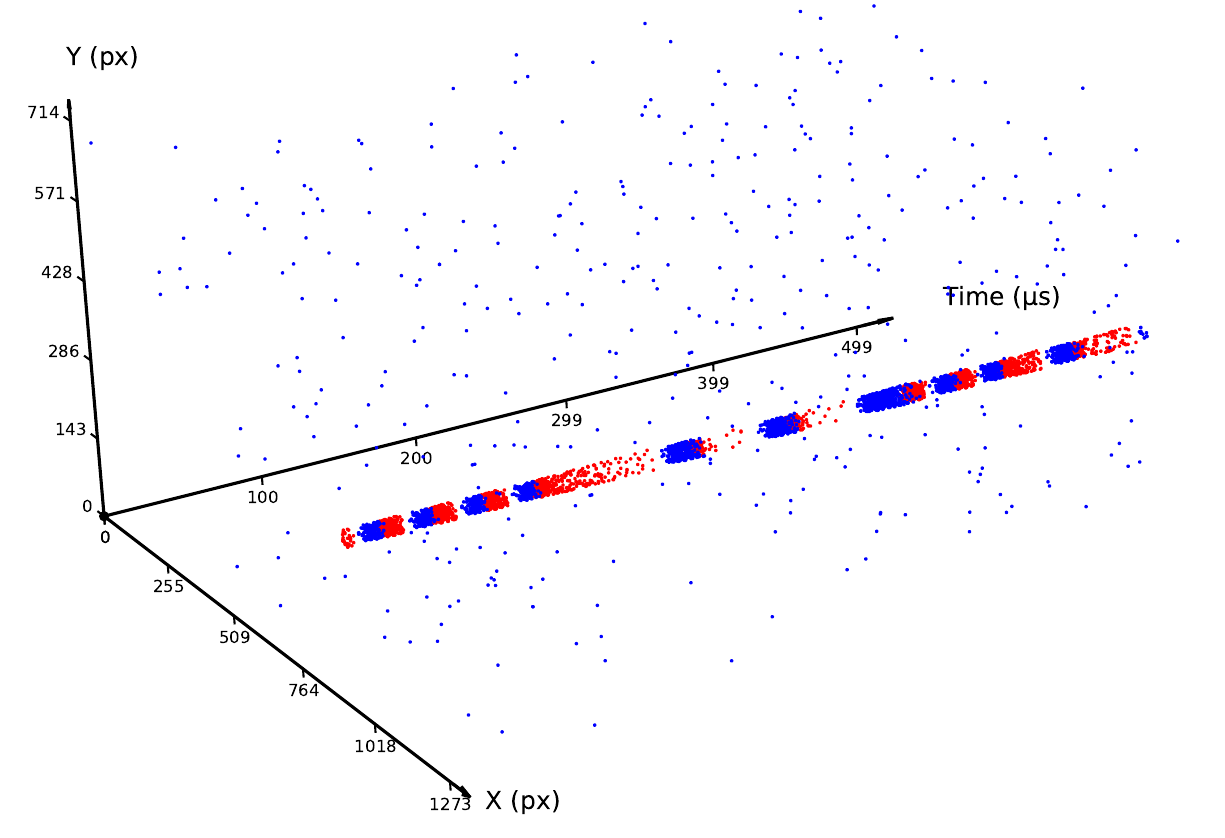}
    \caption{Temporal scatter plot showing \qty{500}{\micro\second} of events generated on the NVS camera by a variable $\SIrange[range-phrase=\text{ to },range-units=single]{20}{64}{\kilo\hertz}$ pulsed beam.
    Scattered dots are events with positive (red) and negative (blue) polarity.}\label{fig:xyt}
\end{figure}

\newpage
\autoref{fig:xyt} illustrates the spatio-temporal representation of a pulsed beam over a \qty{500}{\micro\second} interval. 
The events appear as spaced bands along the time axis, where the spacing between opposite polarities corresponds to the inverse pulse repetition frequency $f_{TX}$.
Since NVSs respond only to changes in light intensity, static background illumination is not visible, and events are generated primarily at signal transitions, where positive and negative polarities correspond to increasing and decreasing intensity, respectively.
The remaining background events represent dark noise and in the presence of platform vibrations, motion-induced contrast changes along the edges of static objects. 
The latter phenomenon is particularly relevant in UAV applications, where spinning propellers or fast moving structural edges within the FoV can act as significant noise sources.
For our laboratory setup described here, even within the most challenging datasets, characterized by daylight illumination and mechanical vibration, the measured background noise peaked at $3$ events/s/pixel ($\approx 3 \times 10^6 \text{ events/s}$ full-frame), whereas the incident laser beacon generated an event rate of $28 \times 10^3$ events/s/pixel ($\approx 7 \times 10^6 \text{ events/s}$ in a tracker).

\begin{figure}[ht]
    \centering
    \includegraphics[width=0.75\linewidth]{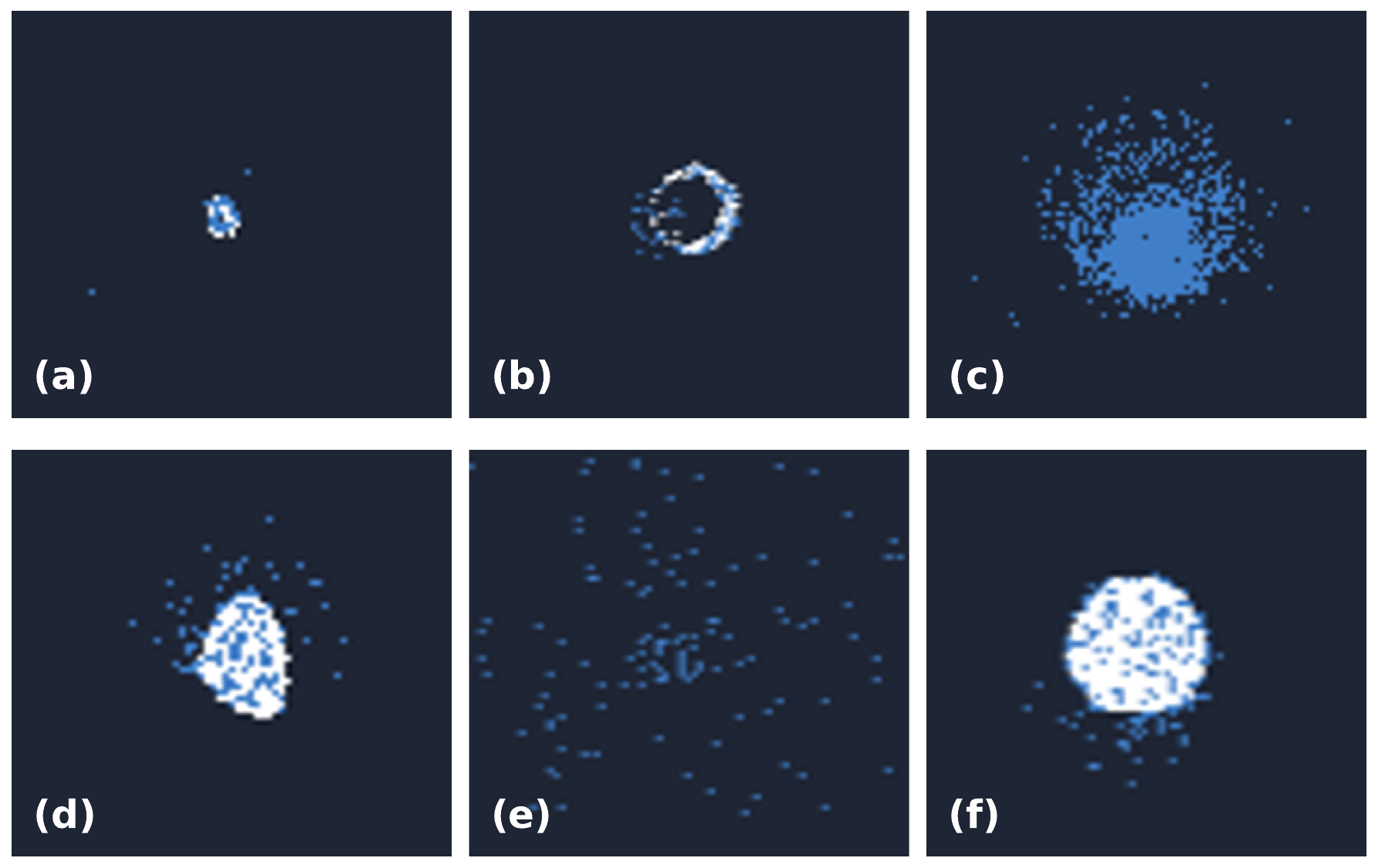}
    \caption{Accumulated events over a \qty{500}{\micro\second} frame, depicting negative (blue) and positive (white) polarities under various conditions: 
    \textbf{(a)} ideal focused beacon; 
    \textbf{(b)} out-of-focus beam; 
    \textbf{(c)} vibration-induced motion blur; 
    \textbf{(d)} coma aberration due to a large angle of arrival; 
    \textbf{(e)} insufficient optical power variation to trigger events; 
    \textbf{(f)} high optical power resulting in an enlarged beam spot.}\label{fig:vary_blobs}
\end{figure}

Besides atmospheric turbulence, which can cause a laser beacon to scintillate, fade, and more generally to
change its size and shape, other challenges arise when searching for features to track. In our experiments, the transmitted beacon is emitted from a single-mode fiber tip (featuring a mode field diameter of \qty{5}{\micro\meter}) and propagates over approximately \qty{2}{\m} of free space.
As illustrated in \autoref{fig:vary_blobs}, where events are accumulated over a \qty{500}{\micro\second} time window, the point spread function can vary significantly depending on the angle of arrival due to optical aberrations from the wide-FoV objective.
Based on the recorded datasets, the analysis in this paper is focused on dynamic only scenarios subjected to mechanical vibrations and signal fading. 
The evaluation utilizes either a single beacon, serving as experimental control, or two beacons to validate multi-target tracking capabilities. 
The dual-beacon architecture serves as a proof of concept for generalized multi-beam ($n \geq 2$) acquisition systems, establishing a baseline for future implementations.

\section{Neuromorphic (event-based) Tracking Algorithms}\label{sec:algorithm}

To process the asynchronous events streams generated by the dynamic scenarios described in the previous section, four distinct tracking algorithms are proposed in this section. All algorithms are asynchronous in that they could process events as they arrive rather than at predefined time intervals.

The \textbf{Kernel-Based Relative Activity Method (KER)} is an implementation of the multi-kernel event-based tracking framework of Lagorce et al.~\cite{lagorceAsynchronousEventBasedMultikernel2015}, where a set of adaptive Gaussian kernels is used to track objects using a NVS. 
In this work, their strategy is adapted to laser beacon tracking by modeling trackers as isotropic Gaussian activity regions. Our choice is motivated by the fact that on the NVS beacons' footprints typically appear as compact, approximately circular intensity patterns whose shape remains largely stable over time, with variations mainly in spatial extent, as previously shown in \autoref{fig:vary_blobs}. The tracker updates the multi-variate standard deviation of the kernel to adapt to the beacon footprint and its mean to track it.

The \textbf{Protocol-Informed Activity Method (PROT)}, outlined in \autoref{alg:pat}, initializes trackers for incoming events to distinguish beacon signals from background noise. Each tracker $i$ maintains a position $(x_c^{(i)},y_c^{(i)})$, radius $r_i$, and activity $a_i$, where activity represents the accumulated event density over one batch period (event density is calculated per batch and accumulated until the end of a deletion interval). 
Events are assigned within a search region defined by the tracker's radius (bounded by a minimum value $r_{\min}$) and a scale factor $s$, and tracker states are updated using the assigned events $E_i$. When an event falls within the search region of multiple trackers, the algorithm uses a greedy approach to assign it to the most active tracker.

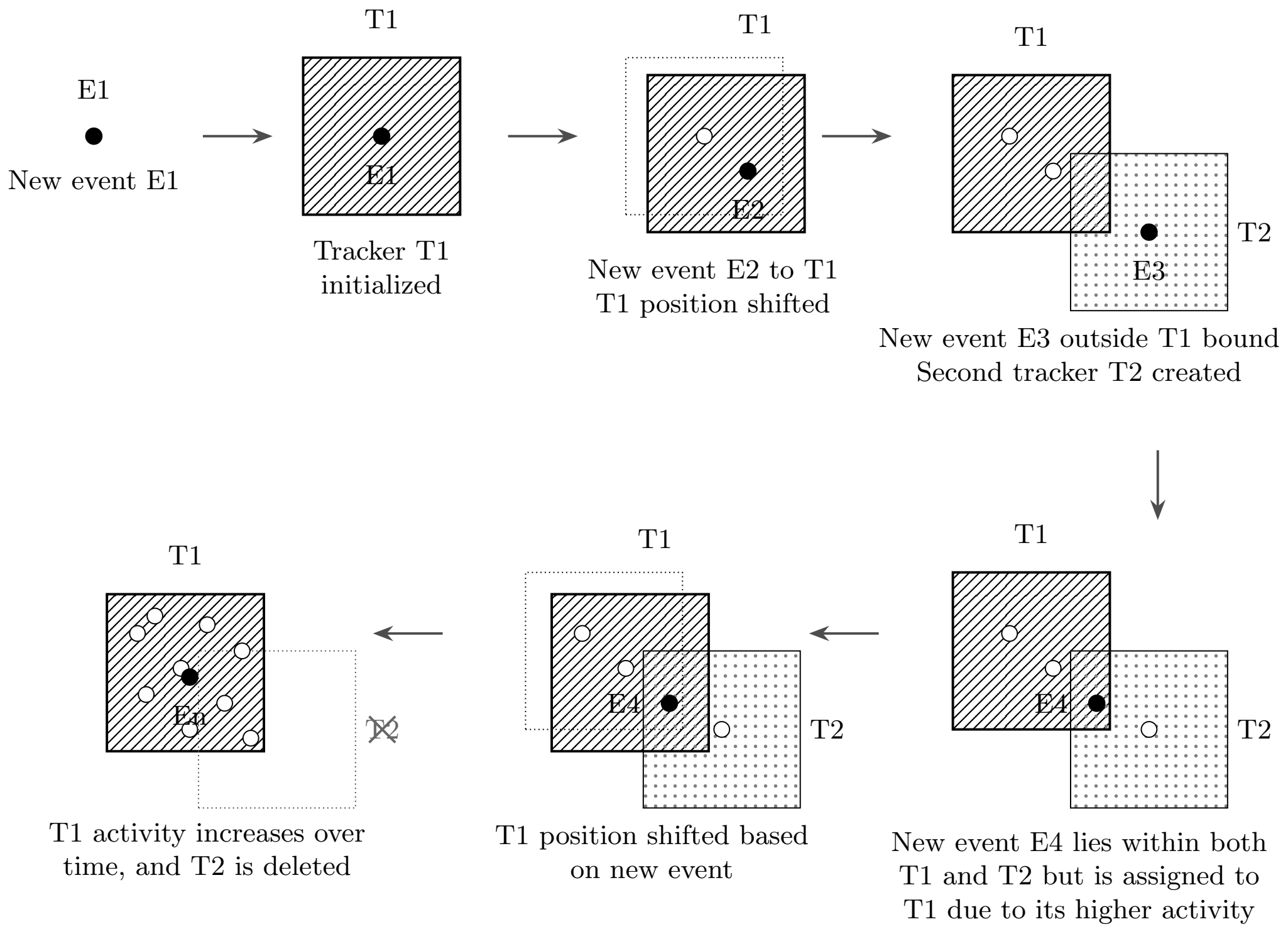
\begin{figure}[H]
    \centering
    \resizebox{0.8\linewidth}{!}{\begin{tikzpicture}[
      node distance = 1.5cm and 1.8cm,
      tracker/.style = {rectangle, draw, thick, minimum size=1.8cm, anchor=center, pattern=north east lines},
      tracker2/.style = {rectangle, draw, minimum size=1.8cm, anchor=center, pattern=dots},
      tracker1/.append style = {pattern color=black!70},
      tracker2/.append style = {pattern color=black!50},
      prev tracker/.style = {tracker, draw=black, thin, densely dotted, fill=none},
      deleted tracker/.style = {tracker2, draw=black!70, thin, densely dotted, fill=none},
      current event/.style = {circle, fill=black, inner sep=2pt},
      prev event/.style = {circle, draw, thin, fill=white, inner sep=1.8pt},
      label/.style = {font=\small},
      below label/.style = {font=\small, align=center},
      flow arrow/.style = {-Stealth, thick, black!70}
    ]

    
    \coordinate (step1) at (0,0);
    \node[current event] (e1_dot) at (step1) {};
    \node[above=0.2cm of e1_dot, label] (e1_lab) {E1};
    \node[below label] at ($(step1) + (0, -0.5)$) {New event E1};

    \draw[flow arrow] (1.25,0) -- (2.05,0);

    \coordinate (step2) at (3.3,0);
    \node[tracker] (t1a) at (step2) {};
    \node[above=0.2cm of t1a.north, label] (t1a_lab) {T1};
    \node[current event] (e1_dot_t1a) at (step2) {}; 
    \node[below=0.1cm of e1_dot_t1a, label] (e1_lab_t1a) {E1};
    \node[below label] at ($(step2) + (0, -1.5)$) {Tracker T1\\initialized};

    \draw[flow arrow] (4.75,0) -- (5.55,0);

    \coordinate (step3) at (7,0);
    
    \node[prev tracker] (t1b_prev) at (step3) {}; 
    \node[above=0.15cm of t1b_prev.north east, label, anchor=south east] (t1b_lab) {T1};
    
    \node[tracker] (t1b) at ($(step3) + (0.25, -0.2)$) {}; 
    
    \node[prev event] (e1_dot_t1b) at (step3) {};
    \node[current event] (e2_dot_t1b) at ($(step3) + (0.5, -0.4)$) {};
    
    \node[below=0.1cm of e2_dot_t1b, label] (e2_lab_t1b) {E2};
    \node[below label] at ($(step3) + (0.1, -1.75)$) {New event E2 to T1\\T1 position shifted};

    \draw[flow arrow] (8.35,0) -- (9.15,0);

    \coordinate (step4) at (10.5, 0);
    
    \node[tracker] (t1c) at ($(step4) + (0.25, -0.2)$) {}; 
    \node[above=0.2cm of t1c.north, label] (t1c_lab) {T1};
    
    \node[tracker2] (t2c) at ($(step4) + (1.6, -1.1)$) {}; 
    \node[label] at ($(t2c.east) + (0.3, 0)$) {T2};
    
    \node[prev event] (e1_dot_t1c) at (step4) {};
    \node[prev event] (e2_dot_t1c) at ($(step4) + (0.5, -0.4)$) {};
    \node[current event] (e3_dot_t2c) at ($(step4) + (1.6, -1.1)$) {};
    
    \node[below=0.1cm of e3_dot_t2c, label] (e3_lab_t2c) {E3};
    \node[below label] at ($(step4) + (0.8, -2.5)$) {New event E3 outside T1 bound\\Second tracker T2 created};

    \draw[flow arrow] (12.2, -3.6) -- (12.2, -4.4);

    
    \coordinate (step5) at (10.5, -5.7);
    
    \node[tracker] (t1d) at ($(step5) + (0.25, -0.2)$) {}; 
    \node[above=0.2cm of t1d.north, label] (t1d_lab) {T1};
    
    \node[tracker2] (t2d) at ($(step5) + (1.6, -1.1)$) {}; 
    \node[label] at ($(t2d.east) + (0.3, 0)$) {T2};
    
    \node[prev event] (e1_dot_t1d) at (step5) {};
    \node[prev event] (e2_dot_t1d) at ($(step5) + (0.5, -0.4)$) {};
    \node[prev event] (e3_dot_t2d) at ($(step5) + (1.6, -1.1)$) {};
    
    \node[current event] (e4_dot_t1d) at ($(step5) + (1.0, -0.8)$) {}; 
    \node[left=0.1cm of e4_dot_t1d, label] (e4_lab_t1d) {E4};
    
    \node[below label] at ($(step5) + (0.8, -2.8)$) {New event E4 lies within both\\T1 and T2 but is assigned to\\T1 due to its higher activity};

    \draw[flow arrow] (9, -5.7) -- (8.2, -5.7);

    \coordinate (step6) at (5.6, -5.7);
    
    \node[prev tracker] (t1e_prev) at ($(step6) + (0.25, -0.2)$) {}; 
    \node[above=0.15cm of t1e_prev.north east, label, anchor=south east] (t1e_lab) {T1};

    \node[tracker] (t1e) at ($(step6) + (0.55, -0.45)$) {}; 
    
    \node[tracker2] (t2e) at ($(step6) + (1.6, -1.1)$) {}; 
    \node[label] at ($(t2e.east) + (0.3, 0)$) {T2};
    
    \node[prev event] (e1_dot_t1e) at (step6) {};
    \node[prev event] (e2_dot_t1e) at ($(step6) + (0.5, -0.4)$) {};
    \node[prev event] (e3_dot_t2e) at ($(step6) + (1.6, -1.1)$) {};
    \node[current event] (e4_dot_t1e) at ($(step6) + (1.0, -0.8)$) {};
    \node[left=0.1cm of e4_dot_t1e, label] (e4_lab_t1e) {E4};
    
    \node[below label] at ($(step6) + (0.8, -2.5)$) {T1 position shifted based\\on new event};

    \draw[flow arrow] (4, -5.7) -- (3.2, -5.7);

    \coordinate (step7) at (0.5, -5.7);
    
    \node[tracker] (t1f) at ($(step7) + (0.55, -0.45)$) {}; 
    \node[above=0.2cm of t1f.north, label] (t1f_lab) {T1};
    
    \node[deleted tracker] (t2f_prev) at ($(step7) + (1.6, -1.1)$) {}; 
    
    \node[prev event] at (step7) {}; 
    \node[prev event] at ($(step7) + (0.5, -0.4)$) {}; 
    \node[prev event] at ($(step7) + (1.0, -0.8)$) {}; 
    \node[prev event] at ($(step7) + (0.2, 0.2)$) {};
    \node[prev event] at ($(step7) + (0.8, 0.1)$) {};
    \node[prev event] at ($(step7) + (1.2, -0.2)$) {};
    \node[prev event] at ($(step7) + (0.1, -0.7)$) {};
    \node[prev event] at ($(step7) + (0.6, -1.1)$) {};
    \node[prev event] at ($(step7) + (1.3, -1.2)$) {};
    
    \node[current event] (en_dot) at ($(step7) + (0.6, -0.5)$) {};
    \node[below=0.1cm of en_dot, label] (en_lab) {En};
    
    \node[below label] at ($(step7) + (0.8, -2.5)$) {T1 activity increases over\\time, and T2 is deleted};
    
    \coordinate (t2f_label_pos) at ($(t2f_prev.east) + (0.3, 0)$);
    \node[label, black!60] at (t2f_label_pos) {T2};
    \draw[thick, black!60] ($(t2f_label_pos) + (-0.15, -0.15)$) -- ($(t2f_label_pos) + (0.15, 0.15)$);
    \draw[thick, black!60] ($(t2f_label_pos) + (-0.15, 0.15)$) -- ($(t2f_label_pos) + (0.15, -0.15)$);

\end{tikzpicture}}
    \caption{Tracker initialization for the Protocol-Informed (PROT) algorithm. Incoming events are processed in continuous batches ($\mathcal{B}_t$) to update or initiate trackers. To discriminate between signal and noise, the system periodically evaluates trackers against an activity threshold ($e_{\min}$) and deletion interval ($T_\text{del}$), deleting any tracker that exhibits insufficient activity ($a_i < e_{\min}$).}\label{fig:prot}
\end{figure}

Unlike the kernel-based method that restricts tracker initialization to a fixed grid in a ghost layer (before promoting it to an active layer), PROT tracker initialization can happen anywhere in the camera frame, as depicted in \autoref{fig:prot}. This approach prevents the missed detections that occur in the KER method when beacons appear outside the initial search regions or exceed the predefined number of kernels.
Protocol knowledge also defines a tracker evaluation interval $T_\text{del}$ based on the minimum modulation frequency $f_{\min}$, the maximum number of consecutive identical transitions $N_{\max}$, and the minimum activity threshold $e_{\min}$. Activity is accumulated over $T_\text{del}$ before low-activity trackers are deleted. Trackers surviving the acquisition time $T_\text{acq}$ are classified as beacon trackers $i$ and retained during temporary inactivity shorter than the fade duration $T_\text{fade}$. 

\begin{algorithm}[t]
\caption{Protocol-Informed Activity Method (PROT)}\label{alg:pat}
\begin{algorithmic}[1]
\Require Event batches $\mathcal{B}_t$, $f_{\min}$, $N_{\max}$, $e_{\min}$, $r_{\min}$, $s$, $T_{\text{acq}}$, $T_{\text{fade}}$
\State $T_{\text{del}} \gets e_{\min} \cdot N_{\max}/f_{\min}$, \quad $t_{\text{del}}^{\text{last}} \gets 0$
\For{each batch $\mathcal{B}_t$ at timestamp $t$}
    \For{each event $e = (x_e, y_e)$}
    \State Assign event $e$ to tracker $i^* \gets \arg\max_{i\in\mathcal{C}} a_i$ if $\mathcal{C} = \{i : |x_e - x_c^{(i)}| \leq s r_i \wedge |y_e - y_c^{(i)}| \leq s r_i\} \neq \emptyset$,
    \Statex \quad \quad \quad else initialize new tracker at $(x_e, y_e)$
    \EndFor
    \For{each tracker $i$}
        \State $E_i \gets$ number of events assigned to tracker $i$
        \State Update tracker $i$ central position $(x_c^{(i)}, y_c^{(i)})$ using the spatial mean of the $E_i$ events
        \State Compute standard deviations $\sigma_{x,i}$ and $\sigma_{y,i}$ of the $E_i$ events
        \State $r_i \gets \max\!\left(\sqrt{\sigma_{x,i}^2 + \sigma_{y,i}^2}, \, r_{\min}\right)$
        \If{$E_i/r_i^2 \geq 1$}
            \State $a_i \gets a_i + E_i/r_i^2$
        \EndIf
    \EndFor
    \If{$t - t_{\text{del}}^{\text{last}} \geq T_{\text{del}}$}
        \State Mark trackers surviving $T_{\text{acq}}$ as beacon trackers $i$
        \State Remove trackers with $a_i < e_{\min}$ except $i$ inactive for $< T_{\text{fade}}$, reset activity of the remaining trackers
        \State $t_{\text{del}}^{\text{last}} \gets t$
    \EndIf
\EndFor

\end{algorithmic}
\end{algorithm}

\begin{figure}[h!]
    \centering
    \begin{tikzpicture}
        \node[anchor=south west, inner sep=0] (image) at (0,0) {\includegraphics[width=0.85\linewidth]{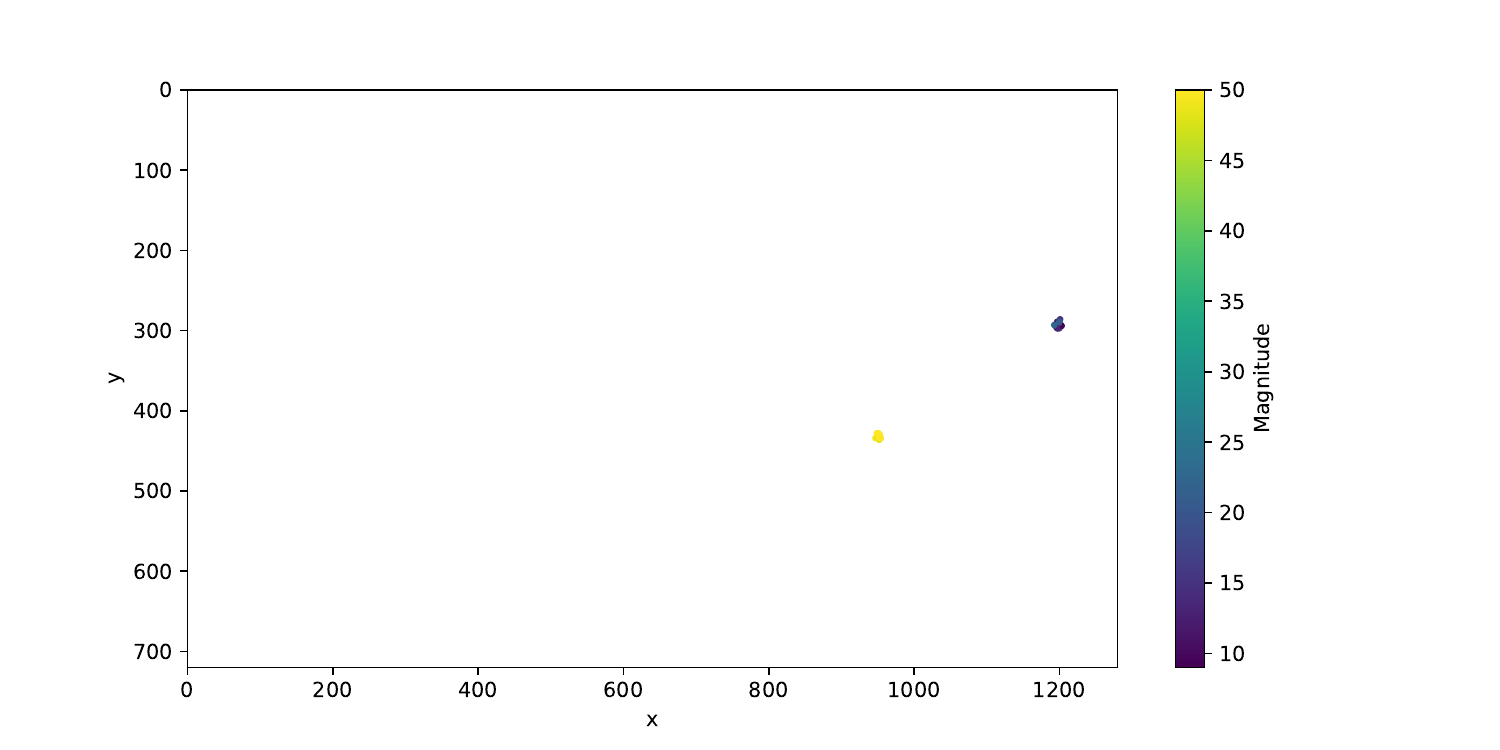}};
        \begin{scope}[x={(image.south east)},y={(image.north west)}]
            \draw[red, thick] (0.695, 0.55) rectangle (0.715, 0.59);
            \draw[red, thick] (0.575, 0.40) rectangle (0.595, 0.44);
        \end{scope}
    \end{tikzpicture}
    \caption{Image of the dominant period $T_p^*$ per pixel for the Histogram Transition Interval (HIST) algorithm. Blue pixels correspond to $ \approx \qty{15}{\micro\second}$ (\qty{65536}{\hertz}), while yellow pixels correspond to $ \approx \qty{50}{\micro\second}$ (\qty{20100}{\hertz}) showing two clusters (red boxes indicate the location of the detected beacons) that correspond to two distinct TX frequencies. }\label{fig:dt_map}
\end{figure}

The \textbf{Histogram Inter-Transition Interval Method (HIST)}, detailed in \autoref{alg:hist_method}, uses inter-transition intervals at each pixel to identify frequencies consistent with the observed rising (or falling) signal edges and localize the corresponding beacons. 
The input edge signal is a pulse train, with discrete temporal resolution of \qty{1}{\micro\second}, where each pulse $k$ at timestamp $t^{(k)}$ corresponds to a polarity transition in a given pixel. 
For each pixel $p=(x_p, y_p)$, consecutive edge timestamps $\{t_{p}^{(k)}\}$ are used to compute inter-transition intervals $\delta t_{p}^{(k)} = t_{p}^{(k+1)} - t_{p}^{(k)}$, from which the pixel's dominant period $T_p^*$ is estimated. Intervals shorter than $\epsilon_{\min}$ are discarded to filter out bursts occurring faster than the maximum candidate frequency.
An example of this dominant period map is shown in \autoref{fig:dt_map}, where distinct spatial clusters corresponding to different modulation frequencies are visible. 
Clusters of pixels are matched to a set of candidate frequencies $\{f_i\}$, each representing a possible tracker $i$, using a number $m$ of bins $\{\mathcal{B}_m\}$ with threshold $\theta_q$ (which defines the minimum count for a valid histogram bin). Within each cluster, the tracker updates the beacon's center position by calculating a weighted centroid based on the period mismatch of each individual pixel (using a Gaussian weighting function with spread $\sigma_q$) within a maximum radius tolerance $r_{\max}$.

\begin{algorithm}[h!]
\caption{Histogram Inter-Transition Interval Method (HIST)}\label{alg:hist_method}
\begin{algorithmic}[1]
\Require Edge events $\mathcal{E}$, candidate frequencies $\{f_i\}$, $m$, $\theta_q$, $\sigma_q$, $r_{\max}$
\State Compute $T_i \gets 1/f_i$; \quad $\epsilon_i \gets \min\!\left(\frac{|T_i-T_{i+1}|}{3}, \, \epsilon_{i-1}\right)$; \quad $\epsilon_{\min} \gets \min_i(\epsilon_i$)
\ForAll{pixels $p = (x_p, y_p)$}
    \State Compute inter-transition intervals $\delta t_{p}^{(k)}$ and discard intervals below $\epsilon_{\min}$
    \State Build histogram of $\delta t_{p}^{(k)}$ and select bin $m^* \gets \min\{m : \text{count}(\mathcal{B}_m) \ge \theta_q\}$
    \State $T_p^* \gets \operatorname{median}\!\left(\delta t_{p}^{(k)} \in \mathcal{B}_{m^*}\right)$
\EndFor
\ForAll{candidate frequencies $f_i$}
    \State $w_{p,i} \gets \exp\!\left(-\frac{(T_p^*-T_i)^2}{2\sigma_q^2}\right)$, \quad $\mathcal{C}_i \gets \{(x_p, y_p, w_{p,i}) : |T_p^*-T_i| < \epsilon_i\}$
    \State Compute magnitude-weighted dominant region center $(x_d^{(i)}, y_d^{(i)})$ from $\mathcal{C}_i$
    \State Estimate magnitude-weighted beacon centroid $(x_c^{(i)}, y_c^{(i)})$ using pixels within $r_{\max}$ of $(x_d^{(i)}, y_d^{(i)})$
\EndFor
\end{algorithmic}
\end{algorithm}

The \textbf{Discrete Fourier Transform Method (DFT)}, summarized in \autoref{alg:dft_beam}, computes the discrete Fourier transform magnitude $M_{p,i}$ of the edge signal $\mathcal{E}$ at each pixel $p=(x_p,y_p)$ for candidate frequencies $\{f_i\}$, identifying the dominant frequency $f_p^*$ and its maximum magnitude $M_p^*$. Here, $N_p$ denotes the total number of edge transitions recorded at pixel $p$ within the batch window $\{  t_p^{(k)} \}^{N_p}_{k=1}$. 
A filtering process based on the magnitude of the DFT is used to suppress weak responses and retain the pixels most likely belonging to a beacon before the tracker centroid estimation. The parameters $M_{\min}$, $q$, $r_{\max}$ define the absolute magnitude threshold, percentile selection, and maximum spatial search range, respectively.

\begin{algorithm}[h!]
\caption{Discrete Fourier Transform Method (DFT)}\label{alg:dft_beam}
\begin{algorithmic}[1]
\Require Edge events $\mathcal{E}$, candidate frequencies $\{f_i\}$, $M_{\min}$, $q$, $r_{\max}$
\ForAll{pixels $p = (x_p, y_p)$}
    \State Compute DFT magnitude for all $f_i$ using edge timestamps $\{t_{p}^{(k)}\}$: 
    \Statex \quad  \quad$M_{p,i} \gets \frac{1}{N_p} \sqrt{\left(\sum_{k=1}^{N_p} \cos(2\pi f_i t_{p}^{(k)})\right)^2 + \left(\sum_{k=1}^{N_p} \sin(2\pi f_i t_{p}^{(k)})\right)^2}$
    \State $f_p^* \gets \arg\max_{f_i} (M_{p,i})$, \quad $M_p^* \gets \max_{f_i} (M_{p,i})$, \quad $w_p = M_p^* $
\EndFor
\ForAll{candidate frequencies $f_i$}
    \State Compute the $q$-th percentile threshold after frequency matching and minimum magnitude filtering: 
    \Statex \quad \quad  $\tau_q \gets \text{percentile}(\{M_p^* : f_p^* = f_i \wedge M_p^* > M_{\min}\}, \, q)$
    \State $\mathcal{C}_i \gets \{(x_p, y_p, w_p) : f_p^* = f_i \wedge M_p^* > \tau_q\}$
    \State Compute magnitude-weighted dominant region center $(x_d^{(i)}, y_d^{(i)})$ from $\mathcal{C}_i$
    \State Estimate magnitude-weighted beacon centroid $(x_c^{(i)}, y_c^{(i)})$ using pixels within $r_{\max}$ of $(x_d^{(i)}, y_d^{(i)})$
\EndFor

\end{algorithmic}
\end{algorithm}

\section{Acquisition and Tracking of FSOC Beacons}\label{sec:results}

The tracking algorithms presented in \autoref{sec:algorithm} are evaluated using a total of 26 datasets recorded following the methodology described in \autoref{sec:dataset}. 
The captured events, distributed across a $1280 \times 720$\,\unit{\px} frame, are processed in fixed \qty{500}{\micro\second} temporal batches. 
Although the algorithms can operate asynchronously (processing single events upon arrival rather than at predefined time intervals), batching is necessary to handle the extra latency introduced by the proprietary data format (Prophesee EVT3.0) and the OpenEB Python HAL library. 
As a result, frequency-based methods have their computation window tied to the batch size, giving them a minimum detectable modulation frequency of $f_{\min} = \qty{20}{\kilo\hertz}$.
All algorithms are implemented in Python 3.12 and executed on a 64-bit system equipped with an Intel Core i7-1365U CPU (\qty{1.8}{\giga\hertz}) and \qty{16}{\giga\byte} of RAM. Computationally intensive operations are accelerated using Numba just-in-time (JIT) compilation.

The acquisition and tracking of incoming beacons are evaluated using the following metrics:
\begin{itemize}[nosep]
    \item \textbf{Latency} [\qty{}{\micro\second}] is defined as the time difference between the occurrence of beacon-related events and the time at which the tracker provides the corresponding centroid position.
    \item \textbf{Temporal Recall} is the ratio ($\in[0, 1]$) 
    of the cumulative duration during which a beacon is associated with a tracker to its total visibility time in the field of view. 
    \item \textbf{False Positive (FP) Rate} [\qty{}{\counts\per\second}] is the number of false detections per second.
    \item \textbf{Identity (ID) Switch Rate} [\qty{}{\counts\per\second}] measures the temporal continuity of beacon tracking. It is defined as the number of times the identity of a tracker assigned to a beacon changes per second.
    \item \textbf{Positional Error} [\qty{}{\px}] is the Root Mean Square Error (RMSE) of tracking residuals from the fitted linear trajectory of a moving beacon.
\end{itemize}
Kalman Filtering (KF) is applied to activity-based methods (KER and PROT), while Particle Filtering (PF) is applied to frequency-based methods (DFT and HIST), to evaluate the effect of temporal filtering on tracking robustness.
\autoref{tab:activity_results} and \autoref{tab:freq_results} present the quantitative comparison of activity-based and frequency-based beacon tracking approaches, respectively. Values represent median values across all recordings, with inter-quartile ranges shown in parentheses and best results are highlighted with a bold font.

\begin{table}[htbp]
\centering
\caption{Performance comparison of activity-based beacon tracking approaches.}
\label{tab:activity_results}
\resizebox{\textwidth}{!}{%
\begin{tabular}{lcccc}
\hline \hline
& \multicolumn{2}{c}{KER} & \multicolumn{2}{c}{PROT} \\
\cline{2-5}
Metrics & without KF & with KF & without KF & with KF \\
\hline
Latency (\si {\micro\second})    & 1400 (1280, 2260) & 1740 (1540, 2610) & \textbf{650 (620, 730)}    & \textbf{650 (620, 730)} \\
Recall          & \textbf{1.000 (1.000, 1.000)} & \textbf{1.000 (1.000, 1.000)} & 0.999 (0.997, 0.999) & 0.999 (0.997, 0.999) \\
FP Rate (\si{\per\second})         & 0.3 (0, 36.5) & 0.2 (0, 36.3) & \textbf{0 (0, 0)} & \textbf{0 (0, 0)} \\
ID Switch Rate (\si{\per\second})     & 0 (0, 0.03) & \textbf{0 (0, 0)} & \textbf{0 (0, 0)} & \textbf{0 (0, 0)} \\
X-RMSE (px)    & \textbf{0.30 (0.10, 0.84)} & 1.09 (0.15, 3.42) & 0.48 (0.14, 1.24) & 0.42 (0.14, 1.24) \\ 
Y-RMSE (px)    & \textbf{0.07 (0.06, 0.23)} & 0.30 (0.06, 0.14) & 0.18 (0.10, 0.24) & 0.18 (0.10, 0.25) \\
\hline \hline
\end{tabular}%
}
\end{table}

\begin{table}[htbp]
    \centering
    \caption{Performance comparison of frequency-based beacon tracking approaches.}
    \label{tab:freq_results}
    \resizebox{\textwidth}{!}{%
    \begin{tabular}{lcccc}
    \hline \hline
    & \multicolumn{2}{c}{DFT} & \multicolumn{2}{c}{HIST} \\
    \cline{2-5}
    Metrics & without PF & with PF & without PF & with PF \\
    \hline    
    Latency (\si {\micro\second})    & 930 (890, 1080)   & 1430 (1340, 2180) & \textbf{880 (830, 1150)}   & 2060 (1710, 4870) \\
    Recall          & 0.965 (0.873, 0.981) & 0.997 (0.943, 0.999) & 0.998 (0.975, 0.999) & \textbf{1.000 (0.994, 1.000)} \\
    FP Rate (\si{\per\second})    & \textbf{0 (0, 0)} & 1.7 (0.5, 4.8) & \textbf{0 (0, 0)} & 0.2 (0.1, 4.1) \\
    ID Switch Rate (\si{\per\second})    & 0 (0, 2.07) & 0 (0, 0.04) & 0 (0, 0.07) & \textbf{0 (0, 0)} \\    
    X-RMSE (px)    & 8.91 (2.16, 24.33) & 3.30 (2.10, 7.86) & \textbf{3.12 (1.26, 5.69)} & 3.22 (1.39, 9.89) \\ 
    Y-RMSE (px)    & 1.60 (1.07, 2.57) & 1.25 (0.67, 2.34) & \textbf{0.79 (0.71, 1.52)} & 1.19 (0.48, 3.14) \\ 
    \hline \hline
    \end{tabular}%
    }
    \end{table} 

\begin{figure}[ht]
    \centering
    \begin{subfigure}[t]{0.49\linewidth}
        \centering
        \includegraphics[width=0.9\linewidth]{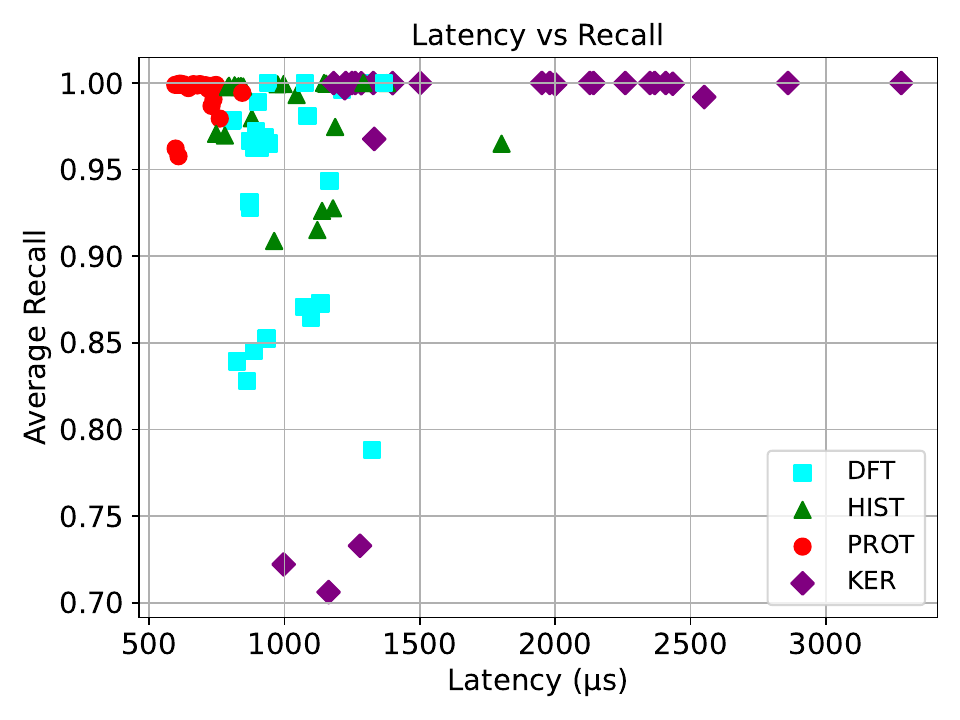}
    \end{subfigure}
    \hfill
    \begin{subfigure}[t]{0.49\linewidth}
        \centering
        \includegraphics[width=0.9\linewidth]{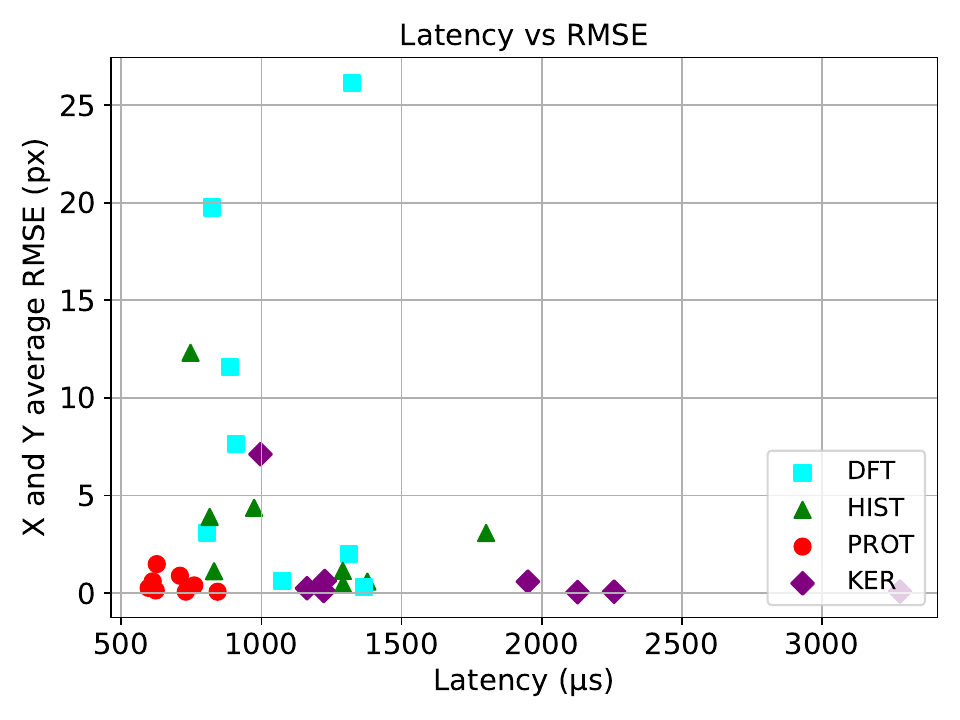}
    \end{subfigure}
    \caption{Scatter plots comparing latency versus tracking accuracy. Marker colors denote the four algorithms. \textbf{(Left)} Accuracy is the median recall across all 26 recordings. \textbf{(Right)} Accuracy is the position error, calculated as the root sum squared along the $x$ and $y$ directions of the RMSE residuals across 8 vibration-free recordings.}
    \label{fig:latency_accuracy}
\end{figure}

The results in \autoref{tab:activity_results} and \autoref{tab:freq_results} are summarized in \autoref{fig:latency_accuracy}. 
PROT achieves the lowest processing latency (\qty{620}{\micro\second} to \qty{730}{\micro\second}) with a median of \qty{650}{\micro\second}, compared to $\approx \qty{900}{\micro\second}$ for DFT and HIST, and $\approx \qty{1400}{\micro\second}$ for KER. Temporal filtering adds computational overhead: PF substantially increases DFT and HIST latency, whereas KF delay on PROT is negligible because its near-zero false positive rate avoids maintaining redundant filters. In contrast, KER's numerous false positives require independent KFs, increasing its median latency to \qty{1740}{\micro\second}.
Atmospheric fading combined with platform vibrations, sensor refractory periods, and variable transmission frequency, causes intervals without generated events. These fades degrade frequency-based methods, which struggle to establish a consistent fundamental frequency when the periodic pattern is disrupted, leading to missed detections, spurious frequency matches, and identity switches between beacons with closely spaced frequencies. While PF improves recall for frequency-based methods by reducing missed detections, it propagates estimation errors and false positives during prolonged fades. Activity-based methods already achieve near-perfect median recall, and almost no identity switches; therefore KF provides limited improvement. Incidentally,  PROT can tolerate long fades through its protocol deletion threshold and maintain without requiring KF prediction under the tested conditions, though KF may be beneficial for stronger turbulence.

\begin{figure}[ht]
    \centering
    \includegraphics[width=0.8\linewidth]{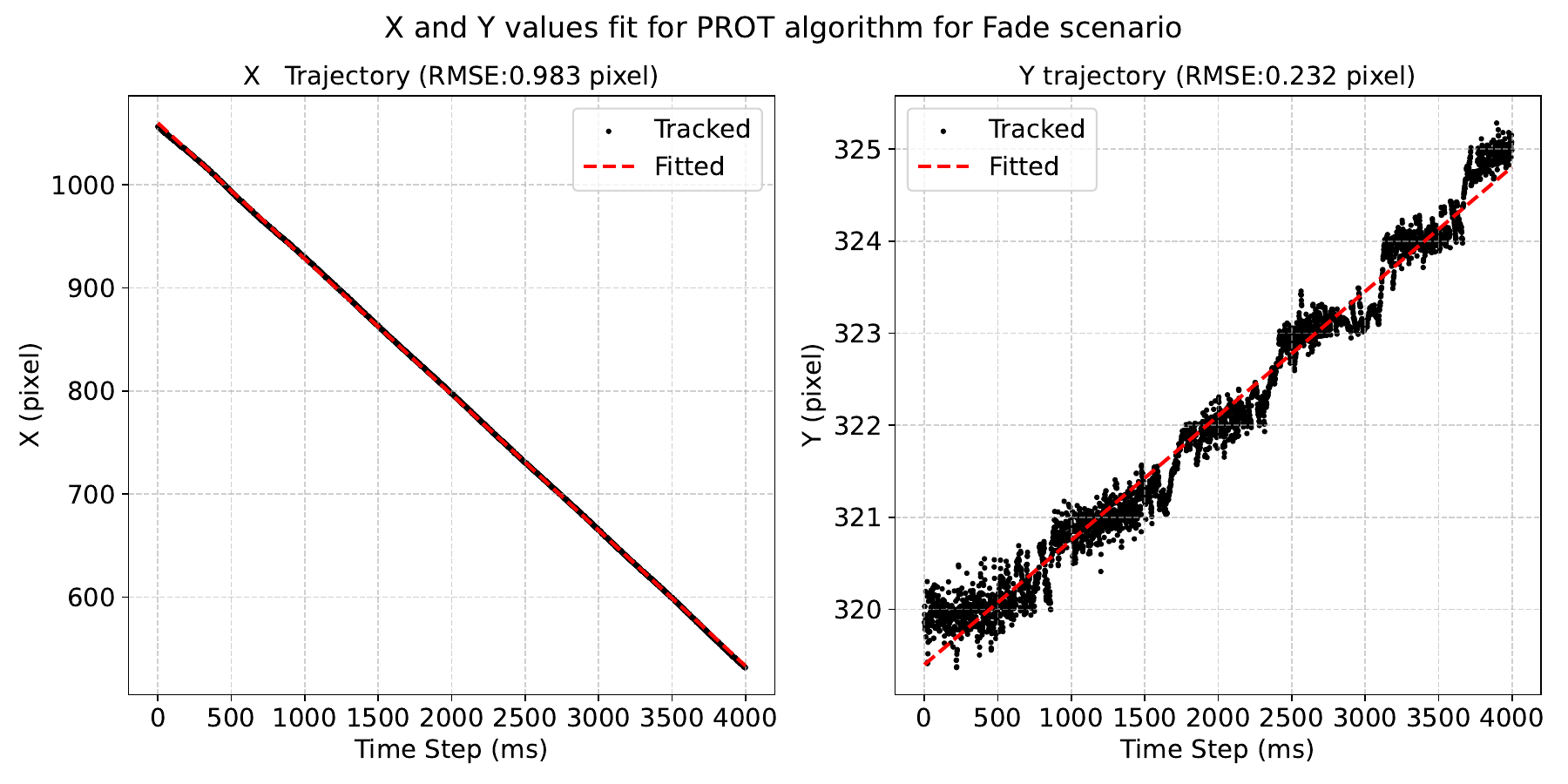} 
    \caption{Example of RMSE calculation by fitting a line on $x$ \textbf{(Left)} and $y$ \textbf{(Right)} tracking directions. This is the trajectory recorded by a dataset without vibrations in the atmospheric fade scenario, where the algorithm PROT is tracking the beacon.}\label{fig:pos_error}
\end{figure}

In recorded datasets without vibration, the precision stage rotates at constant $20 \pm \qty{0.001}{\deg\per\second}$, producing a linear trajectory in the focal plane. Positional accuracy is evaluated by the residual RMSE from a linear fit (\autoref{fig:pos_error}). Under vibration, the limited temporal resolution of the IMU cannot resolve platform motion, so physical oscillations increase the residuals. Activity methods still achieve sub-pixel median accuracy (\autoref{tab:activity_results}), outperforming frequency methods (\autoref{tab:freq_results}), where incorrect frequency association induces larger centroid errors.

\begin{figure}[ht]
    \centering
    \begin{subfigure}[t]{0.49\linewidth}
        \centering
        \includegraphics[width=0.9\linewidth]{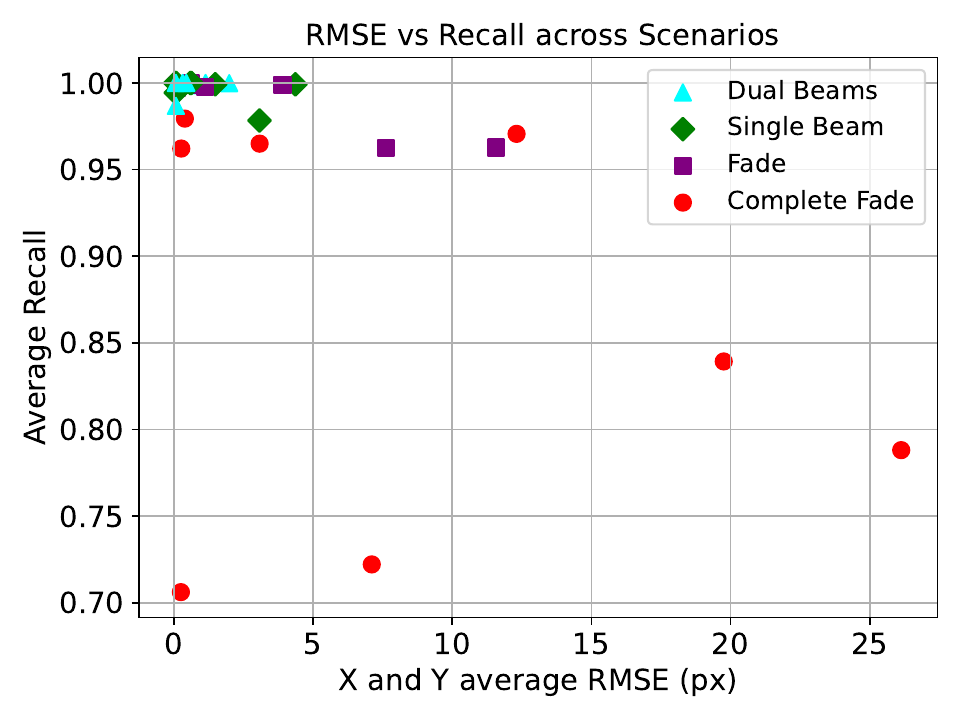}
    \end{subfigure}
    \hfill
    \begin{subfigure}[t]{0.49\linewidth}
        \centering
        \includegraphics[width=0.9\linewidth]{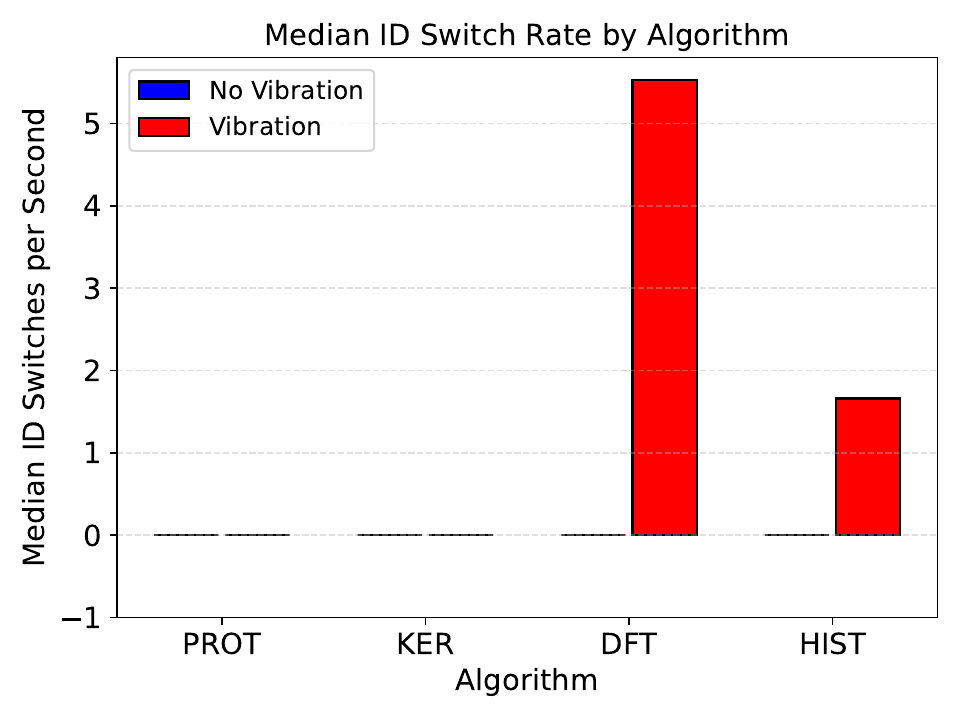}
    \end{subfigure}
    \caption{Evaluation across experimental scenarios \textbf{(Left)} Recall versus positional error across the four scenarios (8 vibration-free recordings, 2 per algorithm). \textbf{(Right)} Effect of platform vibration on median ID switch rate.}
    \label{fig:conditions}
\end{figure}

The outliers visible in \autoref{fig:latency_accuracy}, are explained when differentiating between single beacon, dual beacon, and signal fading datasets, with full results reported in \autoref{sec:app_results}.
In single and dual beacon scenarios, \autoref{fig:conditions}, all methods yield recall above $98\%$ and RMSE below $ \qty{5}{\px}$. Atmospheric fading increases positional error (RMSE up to $\qty{12}{\px}$), while recall remains above $96\%$ for all algorithms. 
The complete fade scenario accounts for the lowest recall values (KER down to $71\%$ and DFT down to $79\%$) as well as the $\qty{26}{\px}$ position error outliner. When optical power drops below sensor sensitivity, KER loses tracking hypotheses and must reacquire them once signal returns, reducing temporal recall. In the absence of signal transitions, DFT locks onto noise frequencies, displacing the estimated centroid.

To measure the extent to which a beacon is successfully acquired and tracked, the recall metric is used. However, recall alone is insufficient for evaluating multi-beacon acquisition; identity preservation must be considered independently. In dual-beacon datasets, platform vibration increase identity switches for frequency-based methods as shown in \autoref{fig:conditions}. When isolating recordings only with platform vibration, frequency-based methods exhibit median ID switch rates of \qty{5.5}{\counts\per\second} for DFT and \qty{1.7}{\counts\per\second} for HIST (whereas non-vibrating datasets bring the global median in \autoref{tab:freq_results} to zero). In contrast, activity-based methods maintain zero median ID switches even under vibration because background clutter generates lower event activity than active beacons.


Currently, the total tracking latency is the sum between the \qty{500}{\micro\second} batching time and the selected algorithm median processing latency. 
Considering the best-performing algorithm (PROT), the tracking bandwidth, i.e. the frequency at which the centroid is updated, is $\Delta B_{\text{t}} = 1 / (\qty{500}{\micro\second} + \qty{650}{\micro\second}) \approx \qty{870}{\Hz}$. 
This bandwidth exceeds both the emulated Greenwood frequency of atmospheric turbulence (assumed $f_G \approx \qty{298.5}{\hertz}$) and UAV platform vibrations harmonics (up to $\qty{283}{\hertz}$, corresponding to a \qty{17000}{\rpm} propeller), while maintaining above $99\%$ temporal recall (excluding complete fade scenarios) and an angular accuracy better than \qty{2}{\milli\radian} with the \qty{3.5}{\mm} wide-FoV objective.

Moreover, provided the beacon's angle of arrival falls within the wide FoV of the neuromorphic camera, the proposed algorithms require a maximum acquisition time of $t_{\text{acq}} = \qty{20}{\milli\second}$ to detect the beacon signal and initiate tracking, thereby significantly extending the effective duration of the communication link.
In scenarios where sub-pixel tracking techniques are not viable, the system's objective lens can be adapted to improve angular resolution over FoV, a trade-off dictated by the sensor's fixed pixel pitch. 

\section{Conclusions and Future Work}\label{sec:conclusion}

In this paper, four event-based tracking algorithms for FSOC beacons were  presented and evaluated under dynamic conditions. 
By utilizing neuromorphic vision sensors, the proposed methods overcome the limitations of traditional frame-based cameras in terms of latency, dynamic range, and motion blur.  
Multi-target tracking (two beacons, generalizable to $n \ge 2$) was experimentally validated in a controlled environment emulating HAP/UAV motion (yaw angular rates up to \qty{20}{\deg\per\s}), three-axis vibrations (up to \qty{1.7}{\m\per\s^2}), and atmospheric scintillation and fading, ultimately demonstrating median recall above 96\% across the benchmark for all algorithms.
The results highlight trade-offs between the approaches. Frequency-based methods exhibit increased identity switches in multi-target scenarios but can distinguish overlapping beacons. Kernel-based methods have better tracking accuracy but in some scenarios can introduce higher median latency. 

Overall, the protocol-informed activity tracker achieves the most balanced performance across all metrics, maintaining near-perfect median recall (above $99\%$), zero median identity switches, and sub-pixel positional error (less than \qty{1}{} pixel) at the lowest median latency of \qty{650}{\micro\second}. If the angle of arrival of a beacon is within the receiver \qty{91.11}{\degree} wide-FoV, the system requires an initial acquisition time of only \qty{20}{\milli\second} and provides a tracking bandwidth of approximately \qty{870}{\hertz} with an angular accuracy better than \qty{2}{\milli\radian}. This demonstrates sufficient capability to track multiple beacons even in the presence of both atmospheric turbulence and HAP/UAV vibrations.

Several limitations apply to this study.
First, the IMX636 sensor operates in the visible spectrum, whereas near-infrared components dominate deployed FSOC systems. As infrared NVSs emerge~\cite{faruhiSWIFTEIWorlds2025}, sub-pixel tracking can compensate for their larger pixel pitch in order to maintain few-milliradian accuracy.  
Second, we emulated atmospheric turbulence effects such as scintillation but omitted wavefront distortion; nevertheless, the tracker radius can be expanded to compute a speckle pattern's global centroid or reduced to treat speckles as independent light sources.

Future work should focus on outdoor testing on longer links distances under real atmospheric turbulence during in-field UAV trials.
Moreover, since batching events introduces a delay, deploying the algorithms on field-programmable gate arrays in order to asynchronously process events and directly control closed-loop pointing, acquisition and tracking should be investigated.
Finally, leveraging the sensor's high temporal resolution (up to $\qty{1}{\MHz}$ bandwidth) for direct data reception via on-off keying~\cite{suminoExperimentalDemonstrationEventbased2025} could enable simultaneous tracking and communication on dynamic platform within a unified receiver architecture. 


\bibliography{references} 

@inproceedings{faruhiSWIFTEIWorlds2025,
	address = {Orlando, United States},
	title = {{SWIFT} {EI}: the world’s first {SWIR} event-based sensor},
	isbn = {978-1-5106-8697-7 978-1-5106-8698-4},
	shorttitle = {{SWIFT} {EI}},
	url = {https://www.spiedigitallibrary.org/conference-proceedings-of-spie/13454/3053313/SWIFT-EI-the-worlds-first-SWIR-event-based-sensor/10.1117/12.3053313.full},
	doi = {10.1117/12.3053313},
	language = {en},
	urldate = {2026-08-06},
	booktitle = {Image {Sensing} {Technologies}: {Materials}, {Devices}, {Systems}, and {Applications} {XII}},
	publisher = {SPIE},
	author = {Faruhi, Rahel and Weinstein, Moshe and Fraenkel, Rami and Jakobson, Claudio and Dobromislin, Roman and Ben Ari, Nimrod and Labilov, Michael and Cohen, Oren and Almog, Ehud and Nitzani, Michal and Karni, Yoram and Markovitz, Tuvy and Amir, Nuriel and Milgrom, Benny Milgrom},
	editor = {Dhar, Nibir K. and Dutta, Achyut K. and Babu, Sachidananda R.},
	month = may,
	year = {2025},
	pages = {8},
}

@article{harrisUseWindowsHarmonic1978,
	title = {On the use of windows for harmonic analysis with the discrete {Fourier} transform},
	volume = {66},
	copyright = {https://ieeexplore.ieee.org/Xplorehelp/downloads/license-information/IEEE.html},
	issn = {0018-9219},
	url = {http://ieeexplore.ieee.org/document/1455106/},
	doi = {10.1109/PROC.1978.10837},
	language = {en},
	number = {1},
	urldate = {2026-07-26},
	journal = {Proceedings of the IEEE},
	author = {Harris, F.J.},
	year = {1978},
	pages = {51--83},
}

@article{welchUseFastFourier1967,
	title = {The use of fast {Fourier} transform for the estimation of power spectra: {A} method based on time averaging over short, modified periodograms},
	volume = {15},
	copyright = {https://ieeexplore.ieee.org/Xplorehelp/downloads/license-information/IEEE.html},
	issn = {0018-9278},
	shorttitle = {The use of fast {Fourier} transform for the estimation of power spectra},
	url = {http://ieeexplore.ieee.org/document/1161901/},
	doi = {10.1109/TAU.1967.1161901},
	language = {en},
	number = {2},
	urldate = {2026-07-26},
	journal = {IEEE Transactions on Audio and Electroacoustics},
	author = {Welch, P.},
	month = jun,
	year = {1967},
	pages = {70--73},
}

@inproceedings{chenDemonstrationBidirectionalCoherent2018,
	address = {San Francisco, United States},
	title = {Demonstration of a bidirectional coherent air-to-ground optical link},
	isbn = {978-1-5106-1533-5 978-1-5106-1534-2},
	url = {https://www.spiedigitallibrary.org/conference-proceedings-of-spie/10524/2292848/Demonstration-of-a-bidirectional-coherent-air-to-ground-optical-link/10.1117/12.2292848.full},
	doi = {10.1117/12.2292848},
	language = {en},
	urldate = {2025-05-16},
	booktitle = {Free-{Space} {Laser} {Communication} and {Atmospheric} {Propagation} {XXX}},
	publisher = {SPIE},
	author = {Chen, Chien-Chung and Jandrain, Dominic and Mak, Gary and Wang, Tony and Grier, Andrew and Malfa, Matt and Booen, Eric M. and Harding, Harvard K. and Xia, Chiyun and Hunwardsen, Matthew T. and Demers, Joseph and Kudinov, Konstantin and Smith, Brett and Sahasrabudhe, Adit and Patawaran, Ferze and Wang, Alan and Zhao, Cen and Leang, Dee and Gin, Jonathan and Lewis, Mark and Nguyen, Danh and Zhang, Boyang and Haque, Farhana and Quirk, Kevin J.},
	editor = {Hemmati, Hamid and Boroson, Don M.},
	month = feb,
	year = {2018},
	pages = {16},
}

@book{andrewsLaserBeamPropagation2005,
	address = {Bellingham, Wash},
	edition = {2. ed},
	series = {{SPIE} {Press} monograph},
	title = {Laser beam propagation through random media},
	isbn = {978-0-8194-5948-0},
	language = {en},
	number = {152},
	publisher = {SPIE Press},
	author = {Andrews, Larry C. and Phillips, Ronald L.},
	year = {2005},
}

@misc{leap1A,
	type = {{TYPE}-{CERTIFICATE} {DATA} {SHEET} ({TCDS})},
	title = {{TYPE}-{CERTIFICATE}  {DATA} {SHEET} {No}. {EASA}.{E}.110  for {LEAP}-{1A} \& {LEAP}-{1C} series engines},
	url = {https://www.easa.europa.eu/en/document-library/type-certificates/engine-cs-e/easae110-leap-1a-leap-1c-series-engines#group-easa-downloads},
	publisher = {EASA},
	author = {{EASA}},
	month = may,
	year = {2026},
	note = {Issue 14},
}

@article{aliPredictingVisibilityLEO1999,
	title = {Predicting the visibility of {LEO} satellites},
	volume = {35},
	copyright = {https://ieeexplore.ieee.org/Xplorehelp/downloads/license-information/IEEE.html},
	issn = {00189251},
	url = {http://ieeexplore.ieee.org/document/805436/},
	doi = {10.1109/7.805436},
	language = {en},
	number = {4},
	urldate = {2026-06-02},
	journal = {IEEE Transactions on Aerospace and Electronic Systems},
	author = {Ali, I. and Al-Dhahir, N. and Hershey, J.E.},
	month = oct,
	year = {1999},
	pages = {1183--1190},
}

@book{hemmatiNearEarthLaserCommunications2021,
	address = {Boca Raton},
	edition = {Second edition},
	series = {Optical science and engineering},
	title = {Near-{Earth} laser communications},
	isbn = {978-1-4987-7740-7 978-0-429-18672-1 978-1-4987-7741-4 978-0-429-53261-0 978-0-429-54731-7},
	doi = {10.1201/9780429186721},
	language = {eng},
	publisher = {CRC Press},
	editor = {Hemmati, Hamid},
	year = {2021},
}

@article{cockram_EventBased_2025b,
	title = {Event-{Based} {Tip}–{Tilt} {Sensing} for {Daytime} {Adaptive} {Optics}},
	volume = {25},
	copyright = {https://creativecommons.org/licenses/by/4.0/legalcode},
	issn = {1530-437X, 1558-1748, 2379-9153},
	url = {https://ieeexplore.ieee.org/document/11220193/},
	doi = {10.1109/JSEN.2025.3624651},
	language = {en},
	number = {24},
	urldate = {2026-05-07},
	journal = {IEEE Sensors Journal},
	author = {Cockram, Monique and Martinez Rey, Noelia},
	month = dec,
	year = {2025},
	pages = {44775--44783},
}

@article{cockram_Eventbased_2025a,
	title = {Event-based detectors for laser guide star tip-tilt sensing},
	volume = {64},
	issn = {0091-3286},
	url = {https://www.spiedigitallibrary.org/journals/optical-engineering/volume-64/issue-04/043102/Event-based-detectors-for-laser-guide-star-tip-tilt-sensing/10.1117/1.OE.64.4.043102.full},
	doi = {10.1117/1.OE.64.4.043102},
	language = {en},
	number = {04},
	urldate = {2026-05-07},
	journal = {Optical Engineering},
	author = {Cockram, Monique and Martinez Rey, Noelia},
	month = apr,
	year = {2025},
}

@misc{suminoExperimentalDemonstrationEventbased2025,
	title = {Experimental {Demonstration} of {Event}-based {Optical} {Camera} {Communication} in {Long}-{Range} {Outdoor} {Environment}},
	url = {http://arxiv.org/abs/2510.14266},
	doi = {10.48550/arXiv.2510.14266},
	publisher = {arXiv},
	author = {Sumino, Miu and Ishii, Mayu and Kaizu, Shun and Hisano, Daisuke and Nakayama, Yu},
	month = oct,
	year = {2025},
	note = {arXiv:2510.14266 [cs]},
}

@article{lichtsteiner128$times$128120DB2008,
	title = {A 128x128 120 {dB} 15 µs {Latency} {Asynchronous} {Temporal} {Contrast} {Vision} {Sensor}},
	volume = {43},
	copyright = {https://ieeexplore.ieee.org/Xplorehelp/downloads/license-information/IEEE.html},
	issn = {0018-9200},
	url = {http://ieeexplore.ieee.org/document/4444573/},
	doi = {10.1109/JSSC.2007.914337},
	language = {en},
	number = {2},
	urldate = {2025-11-24},
	journal = {IEEE Journal of Solid-State Circuits},
	author = {Lichtsteiner, Patrick and Posch, Christoph and Delbruck, Tobi},
	year = {2008},
	pages = {566--576},
}

@article{liAsynchronousVisualInertialOdometry2024,
	title = {Asynchronous {Visual}-{Inertial} {Odometry} for {Event} {Cameras}},
	volume = {XLVIII-4-2024},
	copyright = {https://creativecommons.org/licenses/by/4.0/},
	issn = {2194-9034},
	url = {https://isprs-archives.copernicus.org/articles/XLVIII-4-2024/651/2024/},
	doi = {10.5194/isprs-archives-XLVIII-4-2024-651-2024},
	language = {en},
	urldate = {2026-04-11},
	journal = {The International Archives of the Photogrammetry, Remote Sensing and Spatial Information Sciences},
	author = {Li, Peijing and Yao, Hexiong and Dai, Zhiqiang and Zhu, Xiangwei},
	month = oct,
	year = {2024},
	pages = {651--657},
}

@article{wuEMLFEventbasedFrequency2026,
	title = {E-{MLF}: {Event}-based frequency detection for multiple light sources},
	volume = {193},
	issn = {0030-3992},
	shorttitle = {E-{MLF}},
	url = {https://www.sciencedirect.com/science/article/pii/S0030399225016743},
	doi = {10.1016/j.optlastec.2025.114083},
	urldate = {2025-11-14},
	journal = {Optics \& Laser Technology},
	author = {Wu, Zhiguo and Su, Changqing and Xiong, Bo and Wang, Yu},
	month = jan,
	year = {2026},
	pages = {114083},
}

@article{lagorceAsynchronousEventBasedMultikernel2015,
	title = {Asynchronous {Event}-{Based} {Multikernel} {Algorithm} for {High}-{Speed} {Visual} {Features} {Tracking}},
	volume = {26},
	copyright = {https://ieeexplore.ieee.org/Xplorehelp/downloads/license-information/IEEE.html},
	issn = {2162-237X, 2162-2388},
	url = {https://ieeexplore.ieee.org/document/6899691},
	doi = {10.1109/TNNLS.2014.2352401},
	language = {en},
	number = {8},
	urldate = {2025-11-07},
	journal = {IEEE Transactions on Neural Networks and Learning Systems},
	author = {Lagorce, Xavier and Meyer, Cedric and Ieng, Sio-Hoi and Filliat, David and Benosman, Ryad},
	month = aug,
	year = {2015},
	pages = {1710--1720},
}

@inproceedings{litzenbergerEmbeddedVisionSystem2006,
	address = {Teton National Park, WY, USA},
	title = {Embedded {Vision} {System} for {Real}-{Time} {Object} {Tracking} using an {Asynchronous} {Transient} {Vision} {Sensor}},
	isbn = {978-1-4244-0535-0},
	url = {http://ieeexplore.ieee.org/document/4041053/},
	doi = {10.1109/DSPWS.2006.265448},
	urldate = {2025-11-12},
	booktitle = {2006 {IEEE} 12th {Digital} {Signal} {Processing} {Workshop} \&amp; 4th {IEEE} {Signal} {Processing} {Education} {Workshop}},
	publisher = {IEEE},
	author = {Litzenberger, M. and Posch, C. and Bauer, D. and Belbachir, A.N. and Schon, P. and Kohn, B. and Garn, H.},
	month = sep,
	year = {2006},
	pages = {173--178},
}

@article{cazzatoApplicationDrivenSurveyEventBased2024,
	title = {An {Application}-{Driven} {Survey} on {Event}-{Based} {Neuromorphic} {Computer} {Vision}},
	volume = {15},
	issn = {2078-2489},
	doi = {10.3390/info15080472},
	number = {8},
	journal = {Information},
	publisher = {MDPI AG},
	author = {Cazzato, Dario and Bono, Flavio},
	month = aug,
	year = {2024},
	pages = {472},
}

@article{vonarnimDynamicEventbasedOptical2024,
	title = {Dynamic event-based optical identification and communication},
	volume = {18},
	issn = {1662-5218},
	doi = {10.3389/fnbot.2024.1290965},
	journal = {Frontiers in Neurorobotics},
	publisher = {Frontiers Media SA},
	author = {von Arnim, Axel and Lecomte, Jules and Borras, Naima Elosegui and Woźniak, Stanisław and Pantazi, Angeliki},
	month = feb,
	year = {2024},
}

@inproceedings{censiLowlatencyLocalizationActive2013,
	title = {Low-latency localization by active {LED} markers tracking using a dynamic vision sensor},
	doi = {10.1109/iros.2013.6696456},
	booktitle = {2013 {IEEE}/{RSJ} {International} {Conference} on {Intelligent} {Robots} and {Systems}},
	publisher = {IEEE},
	author = {Censi, Andrea and Strubel, Jonas and Brandli, Christian and Delbruck, Tobi and Scaramuzza, Davide},
	month = nov,
	year = {2013},
	pages = {891--898},
}

@inproceedings{zhuEventBasedVisualInertial2017,
	title = {Event-{Based} {Visual} {Inertial} {Odometry}},
	doi = {10.1109/cvpr.2017.616},
	booktitle = {2017 {IEEE} {Conference} on {Computer} {Vision} and {Pattern} {Recognition} ({CVPR})},
	publisher = {IEEE},
	author = {Zhu, Alex Zihao and Atanasov, Nikolay and Daniilidis, Kostas},
	month = jul,
	year = {2017},
	pages = {5816--5824},
}

@article{iaboniEventBasedMotionCapture2022,
	title = {Event-{Based} {Motion} {Capture} {System} for {Online} {Multi}-{Quadrotor} {Localization} and {Tracking}},
	volume = {22},
	issn = {1424-8220},
	doi = {10.3390/s22093240},
	number = {9},
	journal = {Sensors},
	publisher = {MDPI AG},
	author = {Iaboni, Craig and Lobo, Deepan and Choi, Ji-Won and Abichandani, Pramod},
	month = apr,
	year = {2022},
	pages = {3240},
}

@inproceedings{wangSmartVisualBeacons2022,
	title = {Smart {Visual} {Beacons} with {Asynchronous} {Optical} {Communications} using {Event} {Cameras}},
	doi = {10.1109/iros47612.2022.9982016},
	booktitle = {2022 {IEEE}/{RSJ} {International} {Conference} on {Intelligent} {Robots} and {Systems} ({IROS})},
	publisher = {IEEE},
	author = {Wang, Ziwei and Ng, Yonhon and Henderson, Jack and Mahony, Robert},
	month = oct,
	year = {2022},
}

@inproceedings{jianhongMethodPoseTracking2019,
	title = {Method of pose tracking by event camera using {LED} marker points},
	doi = {10.1109/icemi46757.2019.9101448},
	booktitle = {2019 14th {IEEE} {International} {Conference} on {Electronic} {Measurement} \&amp; {Instruments} ({ICEMI})},
	publisher = {IEEE},
	author = {Jianhong, Xu and Jianwen, Guo and Yan, Wang and Yuheng, Li and Yongdong, Liang and Xiaochang, Cao and Guoliang, Ye},
	month = nov,
	year = {2019},
	pages = {381--386},
}

@inproceedings{mullerMiniatureLowpowerSensor2011,
	title = {A miniature low-power sensor system for real time {2D} visual tracking of {LED} markers},
	doi = {10.1109/robio.2011.6181669},
	booktitle = {2011 {IEEE} {International} {Conference} on {Robotics} and {Biomimetics}},
	publisher = {IEEE},
	author = {Muller, Georg R. and Conradt, Jorg},
	month = dec,
	year = {2011},
	pages = {2429--2434},
}

@article{salahNeuromorphicVisionBasedMeasurement2024,
	title = {A {Neuromorphic} {Vision}-{Based} {Measurement} for {Robust} {Relative} {Localization} in {Future} {Space} {Exploration} {Missions}},
	volume = {73},
	issn = {1557-9662},
	doi = {10.1109/tim.2022.3217513},
	journal = {IEEE Transactions on Instrumentation and Measurement},
	publisher = {Institute of Electrical and Electronics Engineers (IEEE)},
	author = {Salah, Mohammed and Chehadah, Mohammed and Humais, Muhammad and Wahbah, Mohammed and Ayyad, Abdulla and Azzam, Rana and Seneviratne, Lakmal and Zweiri, Yahya},
	year = {2024},
	pages = {1--12},
}

@article{wangAsynchronousBlobTracker2024,
	title = {Asynchronous {Blob} {Tracker} for {Event} {Cameras}},
	volume = {40},
	issn = {1941-0468},
	doi = {10.1109/tro.2024.3454410},
	journal = {IEEE Transactions on Robotics},
	publisher = {Institute of Electrical and Electronics Engineers (IEEE)},
	author = {Wang, Ziwei and Molloy, Timothy and van Goor, Pieter and Mahony, Robert},
	year = {2024},
	pages = {4750--4767},
}
\bibliographystyle{spiebib} 

\appendix    
\newpage
\section{Vibrations from UAV Platforms}\label{sec:app_vibration}

\begin{figure}[H]
    \centering    
    \begin{subfigure}[t]{0.69\textwidth}
        \centering
        \includegraphics[width=\linewidth]{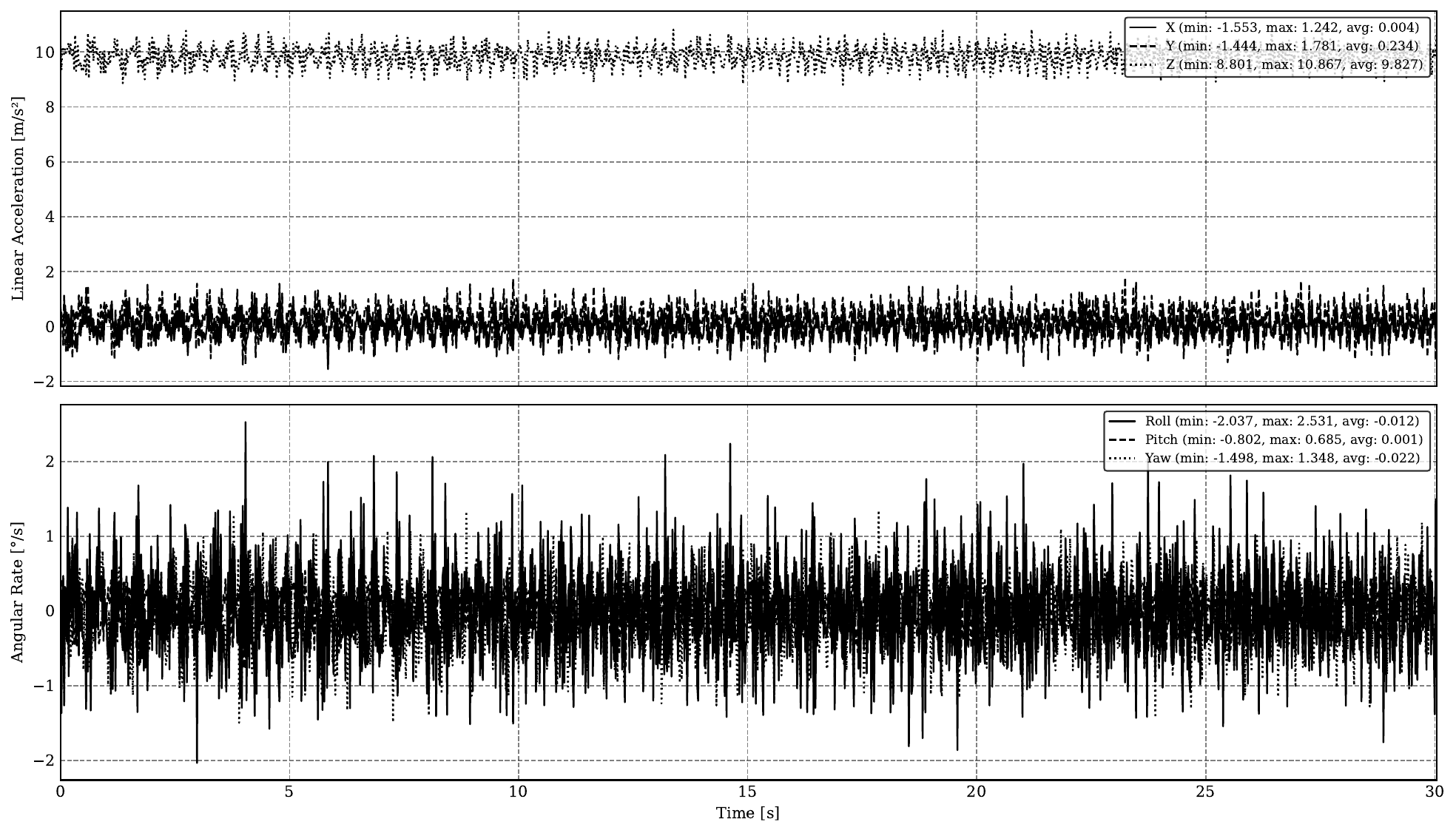}
    \end{subfigure}
    \hfill 
    \begin{subfigure}[t]{0.69\textwidth}
        \centering
        \includegraphics[width=\linewidth]{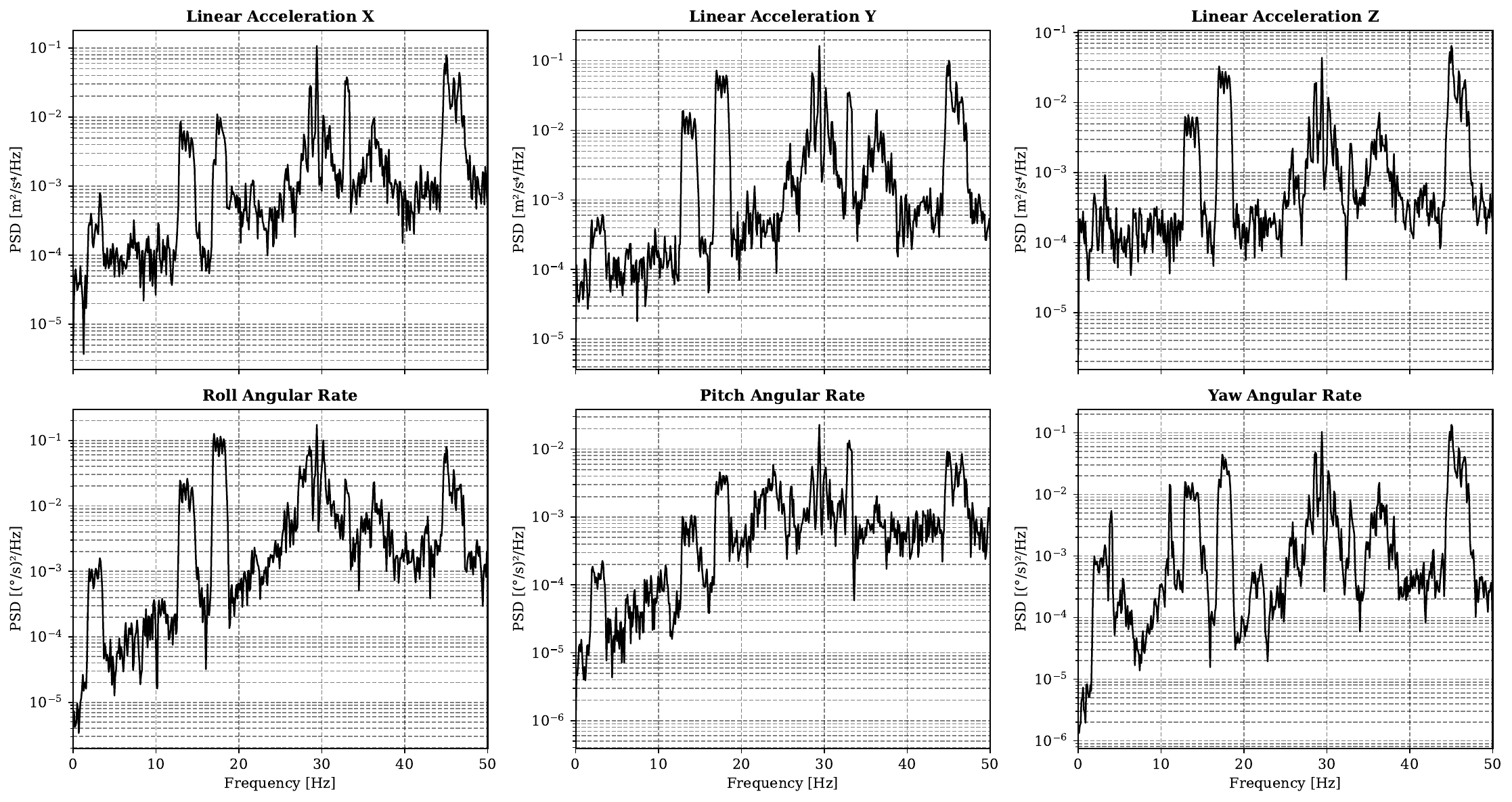}
    \end{subfigure}    
    \caption{Kinematic and vibration analysis of the emulated unmanned aerial vehicle (UAV) environment while hovering. 
    \textbf{(Top)} Inertial data recordings of linear acceleration (including gravity) and angular rate from the Honeywell n380 IMU. For visualization, the series have been down-sampled by a factor of 2.
    \textbf{(Bottom)} Power spectral density of inertial data computed via Welch's method utilizing a Hann window, each data segment length is 1024 samples with 50\% overlap.}
    \label{fig:imu_psd}
\end{figure}

The objective of this experimental setup was to emulate in-flight vibration spectra comparable to those reported in literature, such as the measurements by Chen et al.~\cite{chenDemonstrationBidirectionalCoherent2018}. To achieve this, vibration engines, consisting of an off-centered spinning mass, were employed to emulate vibrations from future unmanned aerial vehicle (UAV) flight test campaigns.
Inertial data, including linear acceleration and angular rate, \autoref{fig:imu_psd} (Top), were acquired utilizing the Honeywell n380 Inertial Measurement Unit (IMU). The time series have a temporal resolution of \qty{10}{\milli\second} (\qty{100}{\hertz} sampling frequency), and are recorded while the swiveling receiver platform is not rotating.

To characterize the vibration profile, the Power Spectral Density (PSD) was estimated using the Welch's method~\cite{welchUseFastFourier1967}, as depicted in \autoref{fig:imu_psd} (Bottom). Prior to spectral analysis, an Hann window \cite{harrisUseWindowsHarmonic1978} is applied to the time-series data and the gravity vector is removed. Welch's method was executed with a segment length of $\num{1024}$ samples and a 50\% overlap. Note that due to the low sampling rate of the IMU, the maximum detectable resonance frequency is significantly lower than the one produce by the vibration engines, nevertheless the inertial data have been proven useful to characterize the angular rate and the maximum vibration amplitude the NVS was subjected.

\newpage
\section{Propagation through atmospheric turbulence}
\label{sec:app_turbulence}

\begin{figure}[H]
  \begin{minipage}[c]{0.6\textwidth}
    \includegraphics[width=\textwidth]{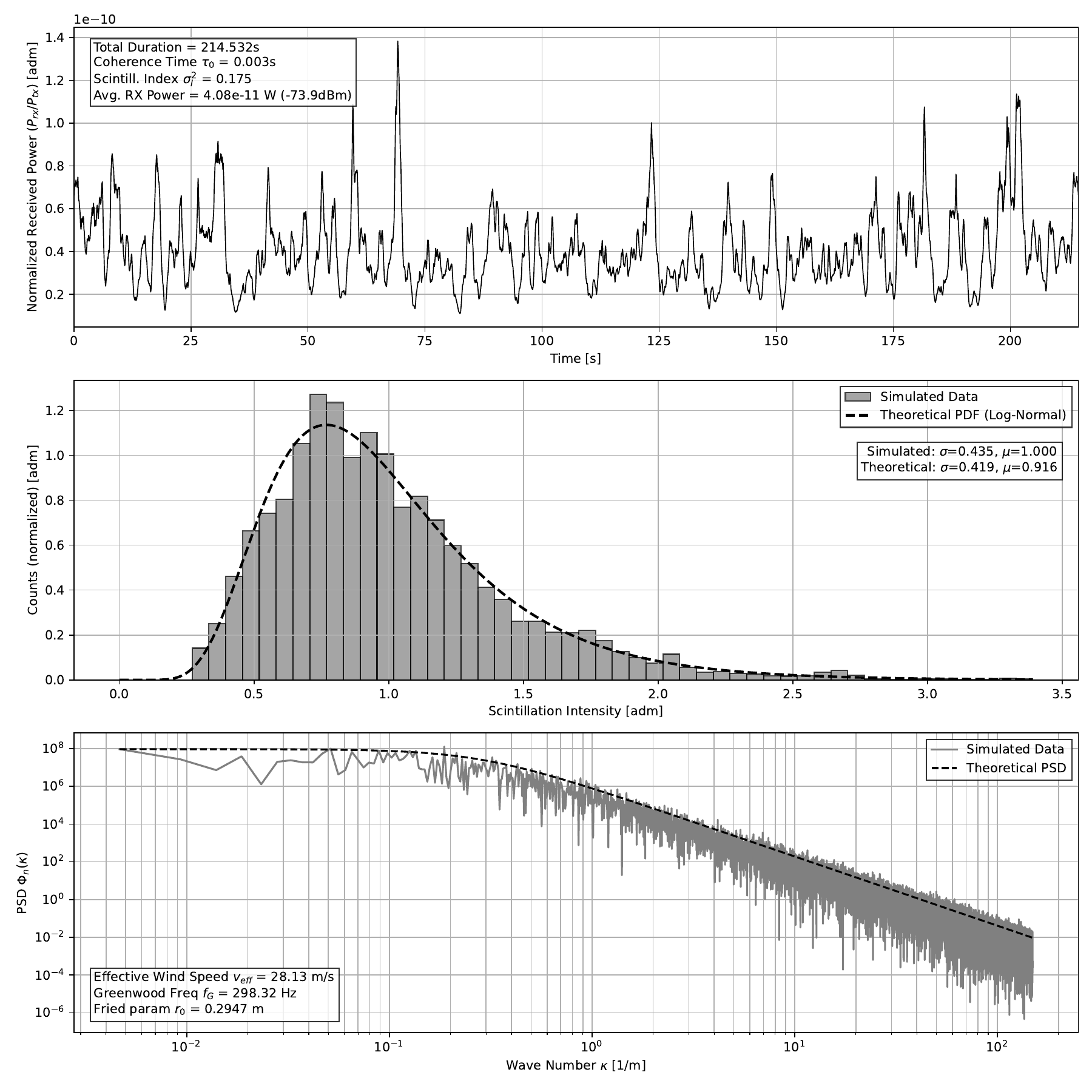}
  \end{minipage}\hfill
  \begin{minipage}[c]{0.37\textwidth}
    \caption{Statistical analysis of the emulated scintillation caused by atmospheric turbulence between two HAPs, at an altitude $h = \qty{10}{\kilo\meter}$ and separated by $L = \qty{20}{\kilo\meter}$. 
    \textbf{(Top)} Normalized ($P_{rx}/P_{tx}$) time series ($N = \num{64000}$ samples) of the received optical power generated over a duration of \qty{214.5}{\second}. The optical link has an average received power of \qty{4.08e-11}{\watt} (\qty{-73.9}{dBm}). 
    \textbf{(Middle)} Probability Density Function (PDF) of the time series. The variance of the emulated data ($\sigma \approx 0.435$) shows agreement with the theoretical ($\sigma \approx 0.419$) Log-Normal distribution for weak turbulence. 
    \textbf{(Bottom)} Power Spectral Density (PSD) of the received power fluctuations, comparing to the frequency-domain shape of the emulated data to the theoretical von Kármán spectrum model. 
    }
    \label{fig:scintillation_diagnostics}
  \end{minipage}
\end{figure}

In order to replicate the effects of atmospheric turbulence on the received power by the NVS, the scintillations statistics described in Andrews and Phillips \cite{andrewsLaserBeamPropagation2005} were emulated. 
Specifically, for a link between two HAPs, a horizontal propagation path at an altitude of $h = \qty{10}{\kilo\meter}$ over a link distance of $L = \qty{20}{\kilo\meter}$ was chosen, utilizing a laser wavelength of $\lambda = \qty{852}{\nano\meter}$  with a plane wave model, transmitting $P_{tx} = \qty{1}{\watt}$. The mean received optical power, $\langle P_{rx} \rangle = \qty{4.08e-11}{\watt}$, is obtained by adding the free space path losses and the atmospheric attenuation across the receiver aperture ($D_{rx} = \qty{19.6}{\milli\meter}$). 
Incidentally, the transverse wind speed, $v_{eff} = \qty{28.13}{\m\per\s}$ is calculated via the Bufton wind model, and the refractive-index structure parameter, $C_n^2$, is described by the Hufnagel-Valley (H-V 5/7) profile.
From these parameters, the theoretical scintillation index is $\sigma_I^2 = 0.175$, the Fried parameter $r_0 = \qty{29.5}{\centi\meter}$, and a time series of $N = \num{64000}$ discrete samples, with time step interval equal to the atmospheric coherence time ($\Delta t = \tau_0 = \qty{3.35}{\milli\second}$) is generated with a two-step Monte Carlo approach to fit the expected amplitude distribution, \autoref{fig:scintillation_diagnostics}. 

For the emulation approach, complex Gaussian noise is generated in the frequency domain. The frequency components are then shaped by filtering them through the theoretical von Kármán spectral model for the refractive-index fluctuations.
The Cumulative Distribution Function (CDF) of each discrete point in the Gaussian sequence is calculated and passed through the inverse CDF of the target physical distribution, i.e. to a Log-Normal distribution for the weak fluctuation regime ($\sigma_I^2 < 1.0$), or a Gamma-Gamma distribution for the strong fluctuation regime ($\sigma_I^2 \ge 1.0$). 
The received (optical) power time series, $P(t)$, is obtained by multiplying this normalized mapped intensity array by the mean received power $\langle P_{rx} \rangle$.
Finally, the so generated intensity scintillation time series was mapped to a required target attenuation via $A_{\text{target}}(t) = A_{\text{min}} - 10 \log_{10}(P(t))$. 
The attenuation time series is converted into an array of voltages with the calibrated response curve of the Variable Optical Absorber (VOA). 
The VOA is driven by an Agilent 33120A arbitrary waveform generator which playback frequency strictly matched to the atmospheric coherence time.

\newpage
\section{ALGORITHM EVALUATION ACROSS DIFFERENT SCENARIOS}
\label{sec:app_results}

\autoref{tab:scenario} presents the evaluation metrics for the four tracking algorithms (DFT, HIST, KER, and PROT) across the experimental scenarios outlined in \autoref{sec:dataset} and presented in \autoref{fig:conditions} from \autoref{sec:results}. All scenarios include baseline platform vibrations and movement:
\begin{itemize}[nosep]
    \item A single beacon tracking (Single Beacon)
    \item Multi-beacons tracking (Dual Beacons)
    \item Multi-beacons tracking in presence of atmospheric fade (Atm. Fade)
    \item Multi-beacons tracking in presence of complete fade (Complete Fade) 
\end{itemize}

\begin{table}[htbp]
    \centering
    \caption{Evaluation metrics for the algorithms proposed in this work across the different scenarios. Values are reported as medians, with inter-quartile ranges shown in parentheses (Q1–Q3).}
    \label{tab:scenario}
    \resizebox{\textwidth}{!}{
    \begin{tabular}{llcccc}
    \hline \hline
    \textbf{Method} & \textbf{Metric} & \textbf{Complete Fade} & \textbf{Dual Beacons} & \textbf{Atm. Fade} & \textbf{Single Beacon} \\
    \midrule
    
    \multirow{4}{*}{DFT}
     & Latency (\si {\milli\second}) & 860 (840, 1090) & 1110 (1080, 1200) & 890 (880, 910) & 870 (810, 970) \\
     & Recall & 0.828 (0.808, 0.834) & 0.954 (0.871, 0.992) & 0.964 (0.955, 0.970) & 0.989 (0.979, 1.000) \\
     & ID Switch Rate (\si{\per\second}) & 0 (0, 0) & 5.53 (0.52, 12.7) & 0 (0, 0) & 0 (0, 0) \\
     & FP rate (\si{\per\second}) & 2.6 (1.4, 10.2) & 0 (0, 0) & 0 (0, 0) & 0 (0, 0) \\
    
    \midrule
    \multirow{4}{*}{HIST}
     & Latency (\si {\milli\second}) & 780 (760, 1290) & 1150 (1060, 1180) & 830 (810, 840) & 1070 (990, 1200) \\
     & Recall & 0.970 (0.967, 0.970) & 0.977 (0.927, 0.998) & 0.998 (0.998, 0.998) & 1.000 (0.999, 1.00) \\
     & ID Switch Rate (\si{\per\second}) & 0 (0, 0.03) & 2.27 (0.31, 7.87) & 0 (0, 0) & 0 (0, 0) \\
     & FP rate (\si{\per\second}) & 27.9 (27.8, 38.0) & 0 (0, 0) & 0 (0, 0) & 0 (0, 0) \\
    
    \midrule
    \multirow{4}{*}{KER}
     & Latency (\si {\milli\second}) & 1160 (1080, 1220) & 2200 (1990, 2400) & 1300 (1240, 1330) & 2610 (2310, 2970) \\
     & Recall & 0.722 (0.714, 0.727) & 1.000 (1.000, 1.000) & 1.000 (1.000, 1.000) & 1.000 (1.000, 1.000) \\
     & ID Switch Rate (\si{\per\second}) & 1.87 (1.46, 1.60) & 0 (0, 0) & 0 (0, 0) & 0 (0, 0) \\
     & FP rate (\si{\per\second}) & 107.2 (106.5, 177.3) & 6.0 (0.3, 26.1) & 0.1 (0, 0.1) & 114.0 (78.1, 140.9) \\
    
    \midrule
    \multirow{4}{*}{PROT}
     & Latency (\si {\milli\second}) & 610 (600, 680) & 720 (690, 740) & 620 (610, 620) & 710 (650, 770) \\
     & Recall & 0.962 (0.960, 0.971) & 0.998 (0.997, 0.999) & 0.999 (0.999, 0.999) & 0.999 (0.998, 0.999) \\
     & ID Switch Rate (\si{\per\second}) & 0.98 (0.97, 0.99) & 0 (0, 0) & 0 (0, 0) & 0 (0, 0) \\
     & FP rate (\si{\per\second}) & 27.8 (26.7, 39.5) & 0 (0, 0) & 0 (0, 0) & 0 (0, 0) \\
    
    \hline \hline
    \end{tabular}
    }
    \end{table}

\end{document}